%% file: 2016_mnras.tex
\documentclass[useAMS,usenatbib]{mn2e}
\usepackage{graphicx}
\input{journals.tex}

\usepackage{times}

\title[Luminous substructure in MG~2016+112]{{The role of luminous substructure in the gravitational lens system MG~2016+112}}
\author[A. More et al.]{A. More$^{1}$\thanks{E-mail: anupreeta@mpifr-bonn.mpg.de, Marie Curie and IMPRS fellow}, J. P. McKean$^{1}$, S. More$^2$, R. W. Porcas$^{1}$, L. V. E. Koopmans$^{3}$ and 
\newauthor M. A. Garrett$^{4,5,6}$\\
$^{1}$Max-Planck-Institut f\"{u}r Radioastronomie, Auf dem H\"{u}gel 69, D-53121 Bonn, Germany\\
$^{2}$Max Planck Institute for Astronomy, Koenigstuhl 17, 69117 Heidelberg, Germany \\
$^{3}$Kapteyn Astronomical Institute, Postbus 800, NL-9700 AV Groningen, the Netherlands \\
$^{4}$Netherlands Institute for Radio Astronomy (ASTRON), Postbus 2, 7990 AA
Dwingeloo, The Netherlands \\
$^{5}$Leiden Observatory, University of Leiden, PB 9513, Leiden 2300 RA, The
Netherlands \\
$^{6}$Centre for Astrophysics and Supercomputing, Swinburne University of Technology, Hawthorn, Victoria 3122, Australia }

\begin{document}

\pagerange{\pageref{firstpage}--\pageref{lastpage}} \pubyear{2002}

\maketitle

\label{firstpage}

\begin{abstract}
MG~2016+112 is a quadruply imaged lens system with two complete images A and B and a pair of merging partial images in region C as seen in the radio. The merging images are found to violate the expected mirror symmetry. This indicates an astrometric anomaly which could only be of gravitational origin and could arise due to substructure in the environment or line-of-sight of the lens galaxy. We present new high resolution multi-frequency VLBI observations at 1.7, 5 and 8.4~GHz. Three new components are detected in the new VLBI imaging of both the lensed images A and B. The expected opposite parity of the lensed images A and B was confirmed due to the detection of non-collinear components. Furthermore, the observed properties of the newly detected components are inconsistent with the predictions of previous mass models. We present new scenarios for the background quasar which are consistent with the new observations. We also investigate the role of the satellite galaxy situated at the same redshift as the main lensing galaxy. Our new mass models demonstrate quantitatively that the satellite galaxy is the primary cause of the astrometric anomaly found in region C. The detected satellite is consistent with the abundance of subhaloes expected in the halo from cold dark matter (CDM) simulations. However, the fraction of the total halo mass in the satellite as computed from lens modeling is found to be higher than that predicted by CDM simulations.
\end{abstract}


\section{introduction}

Cold dark matter (CDM) simulations of hierarchical structure formation predict a large population of small mass subhaloes within the extended dark matter parent haloes of galaxies (e.g., \citealt*{navarro96}; \citealt{moore99}; \citealt*{diemand07}). The total mass fraction of the parent halo that is made up in substructures is thought to be 5--10 per cent and the mass function of the substructures is $ {\rm d}N / {\rm d}M \propto M^{-1.8}$ (e.g., \citealt{diemand07}). Thus, a crucial test of the CDM paradigm is finding these substructures around galaxies like our own Milky Way. Although it appeared that the number of satellites found around our own galaxy, observed as dwarf galaxies, was down by a factor of 10--100 compared to predictions (e.g., \citealt{klypin99,moore99}), in recent years the number of known Milky Way dwarf galaxies has increased (e.g., \citealt{zucker06,belokurov06a,belokurov06b}). However, there still exists a disagreement in the number of observed satellites compared with the predictions from simulations \citep{koposov08}. A possible solution to this `missing satellites' problem is if part of the substructure population is purely dark and devoid of a luminous baryonic component \citep*{bullock01}.

Gravitational lensing offers a powerful method to detect CDM substructures in distant galaxies ($z\la$~1) since the phenomenon is sensitive to matter that is both dark and luminous. The effect of small mass substructures within the lens galaxy is to change the magnifications and/or the positions of the lensed images from what is expected from a globally smooth mass distribution (\citealt{bradac02,bradac04}; \citealt{chiba02,dalal02,metcalf02}; \citealt*{keeton03,keeton05}; \citealt{kochanek04,chen07,mckean07}). 

A statistical analysis of the seven then known radio-loud lens systems\footnote{Gravitational lens systems with a radio-loud lensed source are better for studying flux and astrometric anomalies because radio observations are i) not affected by dust extinction or micro-lensing and ii) provide high resolution astrometry for the lensed images from very long baseline interferometry.} by \citet{dalal02} found that the mass fraction of the substructure required to satisfy the anomalous flux-densities of the lensed images was 0.6--7 per cent of the parent halo (90 per cent confidence level). At first glance, this appeared to be consistent with the expected level of CDM substructure from simulations ($\sim$5--10 per cent). However, \citet{mao04} showed that in the inner part of dark matter haloes, that is, within a few kpc from the centre, which is the region probed by the multiple images of the background source, the expected mass fraction of the CDM substructure is only $\la$~0.5 per cent. Recent CDM simulations have predicted an even smaller mass fraction of $\la$~0.3 per cent \citep{diemand07}.

The discrepancy between the inferred level of CDM substructure from
gravitational lensing and dark matter simulations presents a potential problem
for the CDM paradigm. This has prompted a number of studies into those systems
    with the worst known flux and astrometric anomalies in order to investigate
    their cause. Interestingly, two systems with anomalous lensed image
    flux-densities (CLASS B2045+265 and MG~0414+0534), have been found to
    harbour a small mass (dwarf) companion galaxy to the lens, which when
    included in the lens models, reproduce the observed flux-densities of the
    lensed images \citep{ros00,mckean07}. The gravitational lens system MG~2016+112\footnote{The nomenclature refers to B1950 coordinates. However, all of the observations presented in this paper refer to J2000 coordinates.} \citep{lawrence84}, which is the focus of this paper, shows the most extreme example of the positions of the lensed images differing from the predictions of a globally smooth mass distribution. This astrometric anomaly has also been attributed to a satellite galaxy that is nearby the main lens galaxy \citep{kochanek04,chen07}. In this paper, we present new high resolution radio imaging of the gravitational lens system MG~2016+112 between 1.7 and 8.4~GHz. The purpose of these observations is to test mass models for the lensing galaxy and investigate the contribution, if any, of CDM substructure. 

Our paper is arranged as follows. We first present a review of the previous observations and mass models for MG~2016+112 in Section \ref{2016review}. The new radio observations are presented in Section \ref{obs}. In Section \ref{spec}, the multi-frequency data are used to determine the radio spectral energy distributions of the lensed images. Mass models for the system are investigated in Section \ref{20ma_mod}. We discuss our results in Section \ref{disc} and present our conclusions in Section \ref{conc}. We use the following values of the cosmological parameters $\Omega_{\rm{m}}=0.3$, $\Omega_{\Lambda}=0.7$, $h=0.7$.

\section{MG~2016+112}
\label{2016review}

MG~2016+112 \citep{lawrence84} was the first gravitational lens system to be discovered from the MIT-Green Bank 6-cm survey \citep{bennett86}. The system showed three compact radio components (A, B and C) in VLA imaging (\citealt{lawrence84}) which nearly form a right-angled triangle on the sky. The separation between radio components A and B is $\sim$~3.4~arcsec and the flux-ratio is $S_B/S_A \sim$~1. The third radio component C, which lies $\sim$~2~arcsec south-east of component B, dominates the radio emission from the system; the flux-ratio of components C and A is $S_C/S_A \sim$~3. Observations at a higher angular resolution ($\sim$50~mas beam size) with MERLIN at 5~GHz found components A and B to be compact, however, component C was resolved into essentially two further components, an extended C1 and compact C2, separated by $\sim$~112~mas in an east-west direction \citep{garrett94a}. Very Long Baseline Interferometry (VLBI) observations at 1.7~GHz found extended structure in the components A and B \citep{heflin91} whereas in the European VLBI Network (EVN) observations, component C was resolved into a bead of four components, identified as C11, C12, C13 and C2 \citep{garrett96}. VLBI 5~GHz observations confirmed the presence of extended structure in radio components A, B and C (\citealt{koopmans02b}; hereafter referred to as K02). These data also showed that the components C12 and C13 in region C had similar extended morphologies, consistent with a pair of merging images. The outer pair, C11 and C2, of region C also had a similar morphology, but their surface brightness' were different.

High resolution imaging with the {\it Hubble Space Telescope (HST)} found the optical counterparts of radio components A and B to have strong point-source emission, consistent with a lensed quasar \citep[][CfA-Arizona Space Telescope LEns Survey (CASTLES)]{falco99}. Coincident with radio component C was an extended gravitational arc, which had a redder colour than the quasar emission from A and B \citep*{lawrence93}. Optical spectroscopy showed A and B to be at redshift 3.273 and have similar narrow emission line spectra, typical of a type-2 quasar \citep{lawrence84}. The extended gravitational arc was also found to be at a redshift of 3.273, however, the emission line ratios of the arc were different from those found in lensed images A and B \citep{yamada01}. Finally, the optical imaging also detected a massive elliptical galaxy between the three radio components. The lens, called galaxy D, is at a redshift 1.01 \citep{schneider86} and has a stellar velocity dispersion of 328~km\,s$^{-1}$ \citep{koopmans02a}. The wide separation of components A and B suggested that the lens may be part of a larger structure. Indeed, \citet{soucail01} spectroscopically confirmed an over-density of six galaxies at the same redshift as galaxy D although, subsequent {\it Chandra} observations failed to find evidence for a cluster in the form of extended X-ray emission \citep{chartas01}. Moreover, \citet{clowe01} identified a weak lensing mass peak (3~$\sigma$) $\sim$~64 arcsec northwest of the system.

The similar radio and optical morphologies and spectra of components A and B confirm that they are gravitationally lensed images of the same background source. However, the nature of component C is not so clear, since the radio spectrum is flatter than those of lensed images A and B, the radio morphology appears to be different and the optical emission line ratios are not consistent with those found in the lensed images. A number of mass models have been proposed for MG~2016+112, which have region C being either part of the lens plane (e.g., \citealt{lawrence84,lawrence93}) or part of the background quasar \citep[e.g.,][]{schneider86,langston91} or belonging to both background and foreground (e.g., \citealt{narasimha89,nair97}). The most recent mass model was presented by K02, which had the caustic passing through radio component 2 of the background source (i.e., A2 and B2 in the image plane; see fig. 2 of K02). In this scenario, the radio components 1 and 2 of the background radio source and the optical quasar are doubly imaged, and part of component 2 is quadruply imaged, with the third and fourth images being observed in region C as the merging pair C12 and C13. This picture also explained the extended gravitational arc seen in the optical at region C, since here we are observing the quasar host galaxy without any contamination from the active core. Furthermore, if region C in fact corresponds to two merging images, then we would expect the outer radio components C11 and C2 to have similar properties. However, C2 is found to have a higher flux density. Also, the position of C2 is much farther to the west than expected. A possible solution to this astrometric and flux-ratio anomaly is a nearby satellite galaxy, G1 (\citealt*{kochanek04}; \citealt{chen07}), which is at the same redshift as the lens galaxy D \citep{koopmans02a}.

\section{Radio Observations}
\label{obs}

Radio observations of MG~2016+112 are presented in this section. These include simultaneous MERLIN and global VLBI observations of MG~2016+112 at both 1.7 and 5~GHz, and observations with the High Sensitivity Array (HSA) at 8.4~GHz.

\subsection{MERLIN}
\label{20mer}
\subsubsection{Observations and data reduction}
The MERLIN array was used to observe MG~2016+112 simultaneously with the global VLBI array at 1.7 and 5~GHz. The observations were carried out on 2002 February 25 at 1.7~GHz and on 2001 November 17 at 5~GHz. Both of the experiments had the same observational setup. The data were taken in a single band (referred to as intermediate frequency, IF) of 15~MHz bandwidth and subdivided into 15 channels. The flux calibrator 3C 286 and the point source calibrator B2134+004 were observed for a few minutes each (one$-$two scans). The observations nodded between the lens system MG~2016+112 and the phase calibrator B2029+121. Further observational details can be found in Table \ref{obs_scan}. 
\begin{table}
\begin{center}
\caption{A summary of the MERLIN, VLBI and HSA observations}
\label{obs_scan}
\begin{tabular}{c c c c c}
\hline
Array/     & Total obs. & Integ.   & \multicolumn{2}{c}{Scan length (min)}\\ 
Frequency  & time (h)   & time (s) & Target   &  Phase calibrator\\ \hline
MERLIN    \\ 
1.7~GHz   & 14          &  8   &  3.7   &  1.7  \\ 
5~GHz     & 14          &  8   &  3.7   &  0.8  \\ \hline
VLBI      \\ 
1.7~GHz   & 17          &  2   &  4.5   &  2.4  \\ 
5~GHz     & 17          &  1   &  4.5   &  1.5  \\ \hline
HSA       \\
8.4~GHz   & 7.5         &  1   &  3.5   &  2.3   \\ \hline
\end{tabular}
\end{center}
\end{table}

Using the standard MERLIN data reduction program, the data were first corrected
for any non-closing errors and the amplitude calibration was performed. The
    remainder of the data calibration and imaging was carried out using the
    Astronomical Image Processing Software (\textsc{aips}) provided by the National Radio Astronomical Observatory (NRAO). With the phase referencing technique, the fringe fitting solutions determined for the phase calibrator B2029+121 were applied to calibrate the phase information on MG~2016+112. Furthermore, amplitude and phase self-calibration was performed on the phase calibrator and applied to the target lens system. A phase referenced map of MG~2016+112 was subsequently made using \textsc{imagr}. Using the phase-referenced map as an initial model, phase self-calibration was performed on the lens system. The resulting map was used as a model for a new iteration of calibration of the data and this process was repeated until there was no significant improvement in the produced map. 

\begin{figure*}
\begin{center}
  \includegraphics[width=16cm,height=8cm]{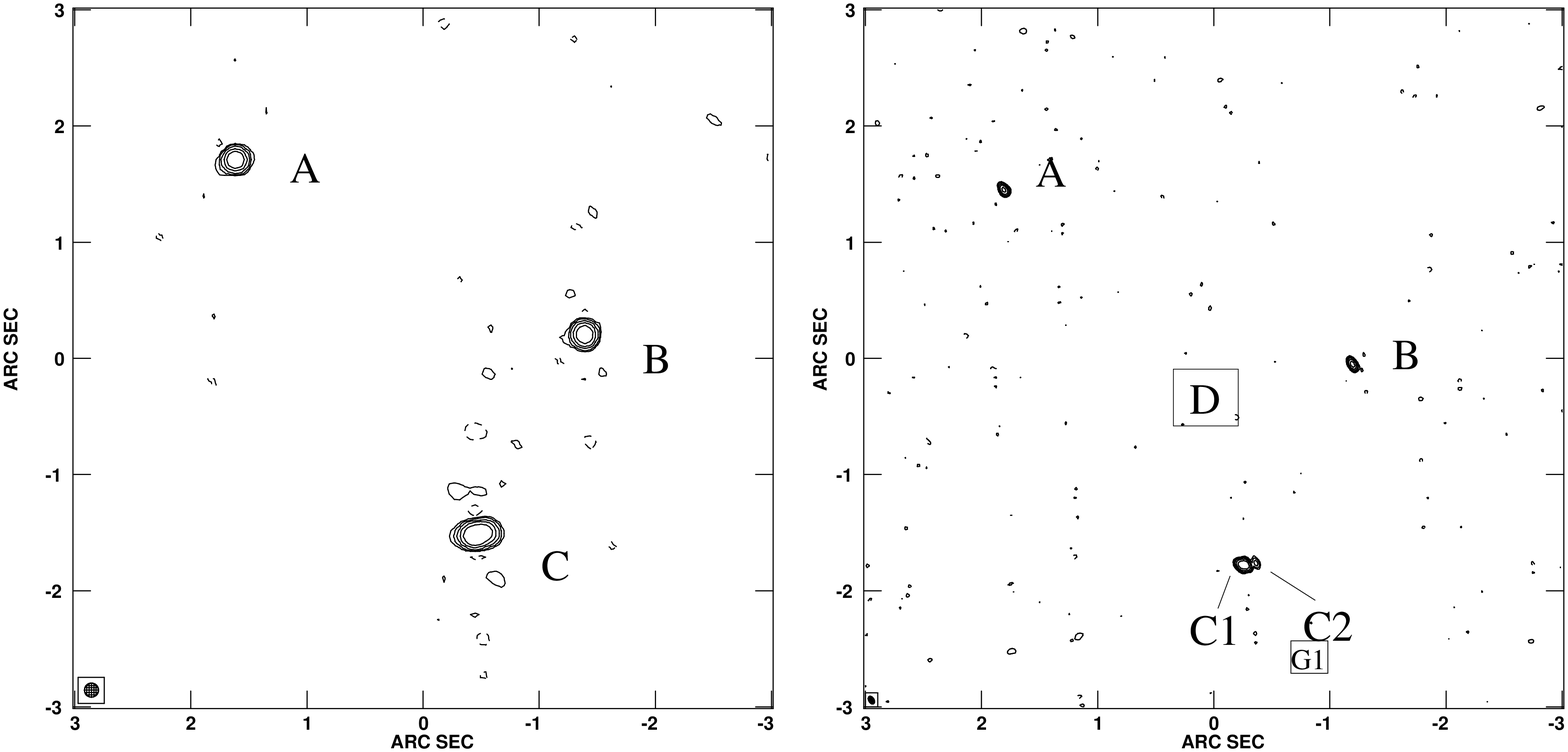}
\caption{{\it Left}: The MERLIN image of MG~2016+112 at 1.7~GHz shows three lensed images (A, B and C). Lensed images A and B are compact whereas region C is slightly extended in the east-west direction. The map is restored with a beam of size 0.12$\times$0.12~arcsec$^2$. {\it Right}: The higher resolution 5~GHz MERLIN image shows that lensed images A and B are still compact, but region C is resolved into a brighter and extended component C1, and a compact component C2 to its west. Furthermore, the squares indicate the positions of galaxies D and G1 found in the optical. The map is restored with a beam of size 0.07$\times$0.04~arcsec$^2$ and a position angle of 31~deg. The root-mean-square (rms) noise in the maps are 0.27~mJy~beam$^{-1}$ at 1.7~GHz and 0.18~mJy~beam$^{-1}$ at 5~GHz. The contour levels are at (-3, 3, 6, 12, 24, 48)~$\times$ the rms noise in the respective maps. North is up and east is to the left.}
\label{mer_186}
 \end{center}
\end{figure*}

\begin{table}
\begin{center}
\caption{The positions, peak surface brightness and total flux densities of MG~2016+112 from the MERLIN observations at 1.7~GHz. The `--' implies that the integrated flux density could not be determined.}
\label{tamer1.7}
\begin {tabular}{l r r c c}
\hline
  Comp. & RA  & Dec   &  $I_{peak}$ & $S_{total}$ \\ 
   &  (mas) & (mas)  &  (mJy~beam$^{-1}$) & (mJy) \\ \hline
  A     &       0$\pm$0.5  &       0$\pm$0.5   &  34.6$\pm$0.3  & 34.9$\pm$0.5 \\ 
  B     & $-$3005.8$\pm$0.5  & $-$1503.9$\pm$0.5   &  34.8$\pm$0.3  & 35.6$\pm$0.5 \\ 
  C1a    & $-$2045.4$\pm$0.9  & $-$3246.2$\pm$0.9   &  13.4$\pm$0.3  & --~~~~~ \\ 
  C1b    & $-$2093.8$\pm$1.4  & $-$3221.5$\pm$1.4   &  42.9$\pm$0.3  & 62.8$\pm$0.6 \\ 
 \hline
\end {tabular}
\end{center}
\end{table}

\subsubsection{Results}
Fig.~\ref{mer_186} shows the MERLIN maps of the lensed images of MG~2016+112 at 1.7 and 5~GHz. The weighting of the {\it uv}-data can be chosen between natural and uniform to give either better resolution or sensitivity at the cost of the other. It is possible however, to optimise the weighting by setting the parameter \textsc{robust} to 0 in \textsc{imagr}. Gaussian model fitting of the components in the image plane was done with the task \textsc{jmfit}. The results are presented in Tables \ref{tamer1.7} and \ref{tamer5}. Lensed images A and B are found to be compact, and were each fitted by a single Gaussian model component at both frequencies. The slightly extended component C was fitted with two Gaussian components at 1.7~GHz and with three at 5~GHz. Note that the components of C1 that were detected at 1.7 and 5~GHz may not be the same. These components were identified by comparing with the model fitting of the high resolution VLBI images (see Section \ref{glo_vlbi}). The relative flux densities of images B, C1 (C1a+C1b) and C2 with respect to image A for the MERLIN 5~GHz imaging are consistent with previous measurements by \citet{garrett94a}. However, the absolute flux densities of the lensed images in the new observations are less by 23 per cent which might be due to errors in the amplitude calibration of either the previous or new data sets. 
\begin{table}
\begin{center}
\caption{The positions, peak surface brightness and total flux densities of MG~2016+112 from the MERLIN observations at 5~GHz. }
\label{tamer5}
\begin {tabular}{l r r c c}
\hline
  Comp. & RA  & Dec   &  $I_{peak}$ & $S_{total}$ \\ 
   &  (mas) & (mas)  &  (mJy~beam$^{-1}$) & (mJy) \\ \hline
  A     &     0.0$\pm$0.5  &     0.0$\pm$0.5   & 11.7$\pm$0.2  & 12.1$\pm$0.3  \\ 
  B     &  $-$3003.2$\pm$0.5  &  $-$1502.0$\pm$0.5   & 13.1$\pm$0.2  & 13.1$\pm$0.3  \\ 
  C1a   &  $-$2048.3$\pm$0.5  &  $-$3230.5$\pm$0.5   & 15.4$\pm$0.2  & 20.0$\pm$0.4  \\ 
  C1b   &  $-$2090.8$\pm$0.5  &  $-$3225.6$\pm$0.5   & 13.1$\pm$0.2  & 13.2$\pm$0.3  \\ 
  C2    &  $-$2171.5$\pm$1.5  &  $-$3212.8$\pm$1.5   &  3.1$\pm$0.2  &  3.0$\pm$0.3  \\ 

 \hline
\end {tabular}
\end{center}
\end{table}

\subsection{Global VLBI} 
\label{glo_vlbi}
\subsubsection{Observations and data reduction}
Earlier 1.7~GHz EVN observations \citep{garrett96} revealed fine structure in region C. With a high resolution spectral analysis, the predictions of the spectra of the components in region C from some of the mass models \citep[e.g,][]{nair97} could be tested. Therefore, high resolution global VLBI observations of MG~2016+112 were undertaken at 1.7 and 5~GHz on 2002 February 25 and 2001 November 17, respectively. 

Since the lensed images of MG~2016+112 have a low flux density, phase referenced observations are vital in determining the phase corrections on all baselines. A strong source within 2~deg of the lens system (B2029+121 with a total flux density $\sim$~0.9~Jy), was used as the phase calibrator for these observations. The scan lengths for each observation of MG~2016+112 and the calibrator B2029+121, the correlator integration time and the total time of the observations are listed in Table \ref{obs_scan}. The data were taken in four IFs at 1.7~GHz and in two IFs at 5~GHz. Each IF had a bandwidth of 8~MHz and was further divided into 16 channels.

The antennas used for the 1.7~GHz observations were Effelsberg (Eb), Jodrell Bank (Jb), Medicina (Mc), Onsala (On), Torun (Tr) and the 10 antennas of the VLBA. In addition, the phased VLA (Y) and the 70~m dishes of Robledo and Goldstone were included for better sensitivity. Arecibo (Ar) and Westerbork (Wb) were also used, however, these data were lost at the VLBA correlator. At 5~GHz, all of the VLBA antennas and Eb, Jb, Mc, On, Tr, Ar, Y and Wb were used. 

The data were calibrated in the standard manner within \textsc{aips}. The data were also corrected for the change in the parallactic angle to account for the apparent change in the position angle as the source moves across the sky. Subsequently, the {\it a priori} amplitude calibration was performed using the system temperatures of each antenna. To determine the system temperature of the phased VLA during the observations, the amplitude level of the phase calibrator was set to the known values obtained from the list of calibrators at different frequencies. 

The data for the calibrator were fringe fitted on the longest scan to determine the phase, delay and rate solutions for a single IF initially, and then for all of the IFs combined, which were then applied to the whole data set. Since averaging the data across the bandwidth can result in loss of amplitude and phase information in the wide-field maps, the data channels were not averaged, in spite of the large number of visibilities. The phase calibrator was mapped by performing phase self-calibration and then both phase and amplitude calibration. The phase and amplitude corrections from the best calibrated map of B2029+121 were then applied to the target. Subsequently, the data on the lens system were phase self-calibrated with solution intervals of 3 and 2~min for the 1.7 and 5~GHz datasets, respectively. 

For MG~2016+112, the lensed sources within the field have a large angular separation (e.g., a few arcseconds). Hence, multiple small fields centred on the lensed images were mapped, that is, three windows centred at the position of A, B and region C were \textsc{clean}ed while mapping MG~2016+112 with \textsc{imagr}. At 1.7~GHz, the weighting was chosen to produce an optimal balance between resolution and sensitivity, whereas at 5~GHz the resolution was good enough, so only the sensitivity had to be up-weighted, hence, natural weighting was applied. Similar to the technique used for the reduction of MERLIN data, self-calibration and mapping were iteratively carried out until no significant improvement was found in the maps.

\subsubsection{The lensed images A and B}

\begin{figure*}
\begin{center}
  \includegraphics[scale=0.3]{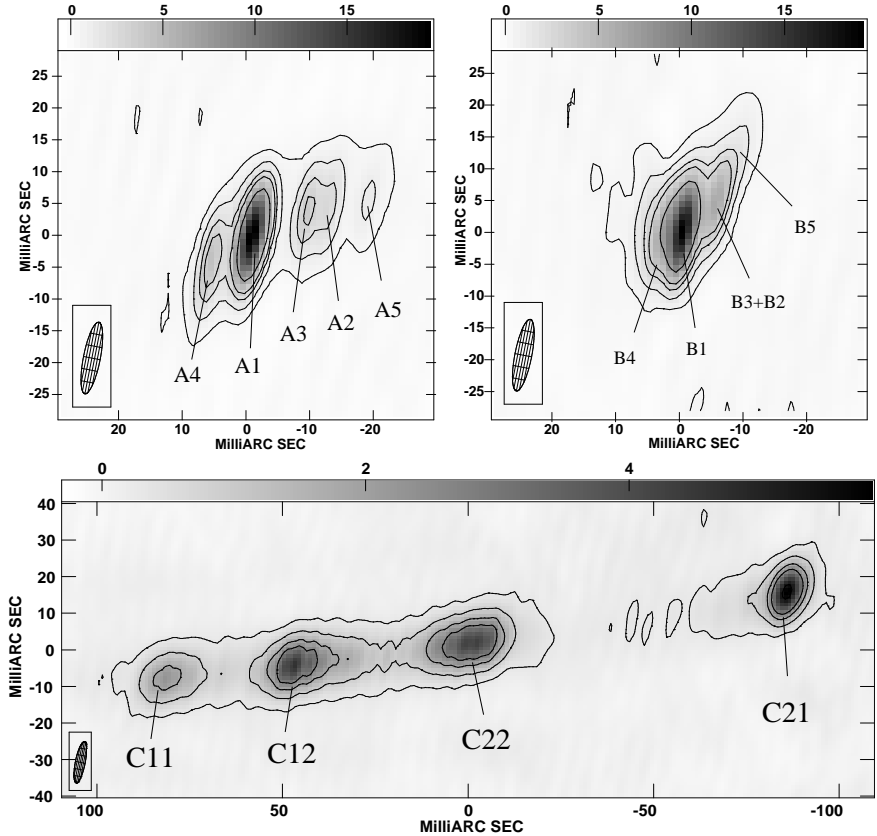}
  \caption{The global VLBI maps of lensed images A and B, and region C at 1.7~GHz with a weighting between uniform and natural. The rms noise in the maps is 0.08~mJy~beam$^{-1}$ and the contours are (-3, 3, 6, 12, 24, 48)~$\times$ the rms noise in the map. The size of the restoring beam is 11.1$\times$2.6~mas$^2$ and the position angle is $-$10.6 deg. The grey-scales are in mJy~beam$^{-1}$. North is up and east is left. }
\label{glo_1.7}
 \end{center}
\end{figure*}

\begin{figure*}
\begin{center}
  \includegraphics[scale=0.3]{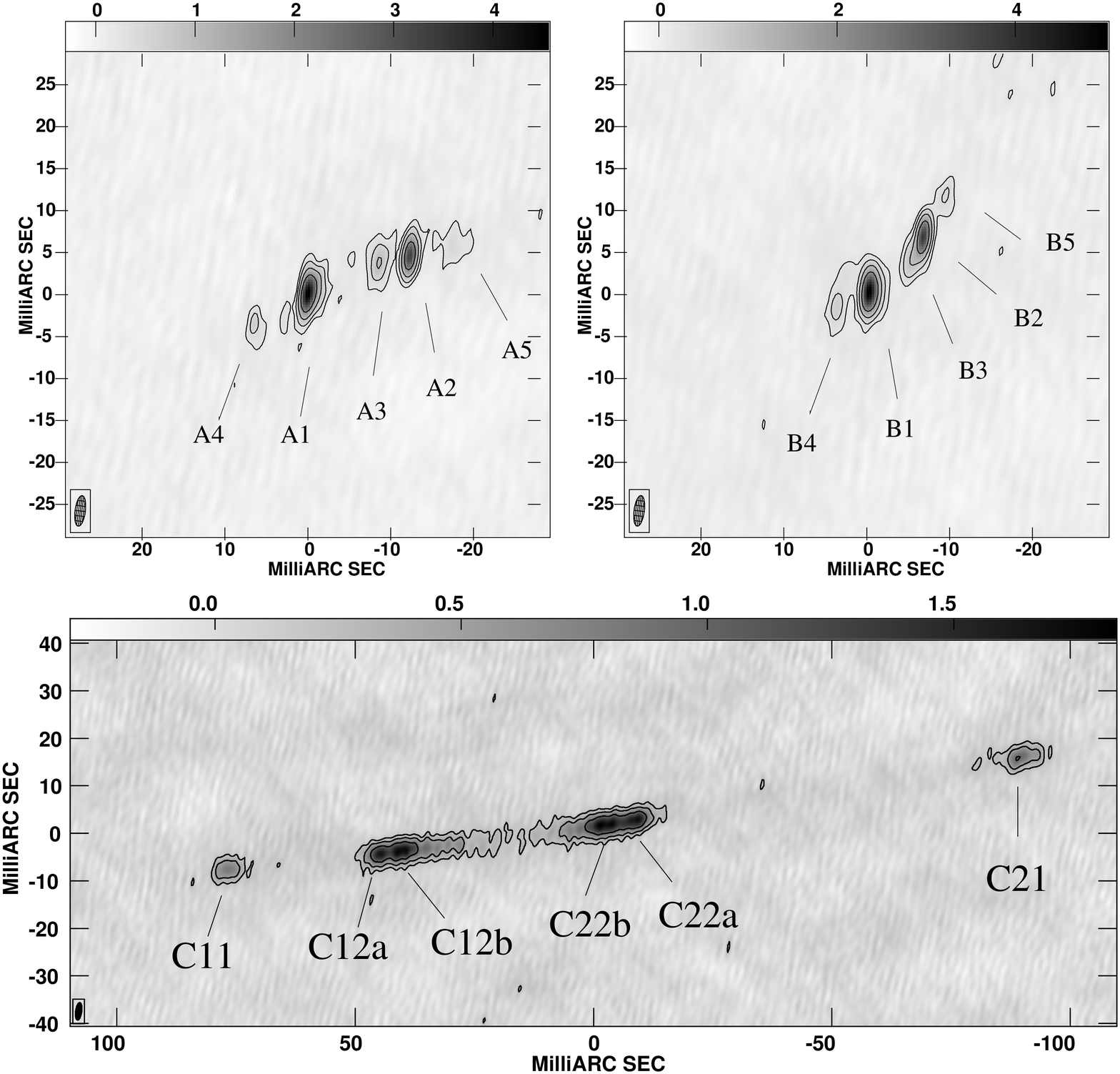}
\caption{The global VLBI maps of lensed images A and B, and region C at 5~GHz with natural weighting. The rms noise in the maps is 0.06~mJy~beam$^{-1}$ and the contours are ($-$4, 4, 8, 16, 32)~$\times$ the rms noise in the map. The size of the restoring beam is 3.7$\times$1.2~mas$^2$ and the position angle is $-$7.54 deg. The grey-scales are in mJy~beam$^{-1}$. North is up and east is left.}
\label{glo_5}
 \end{center}
\end{figure*}

\begin{figure*}
\begin{center}
  \includegraphics[scale=0.3]{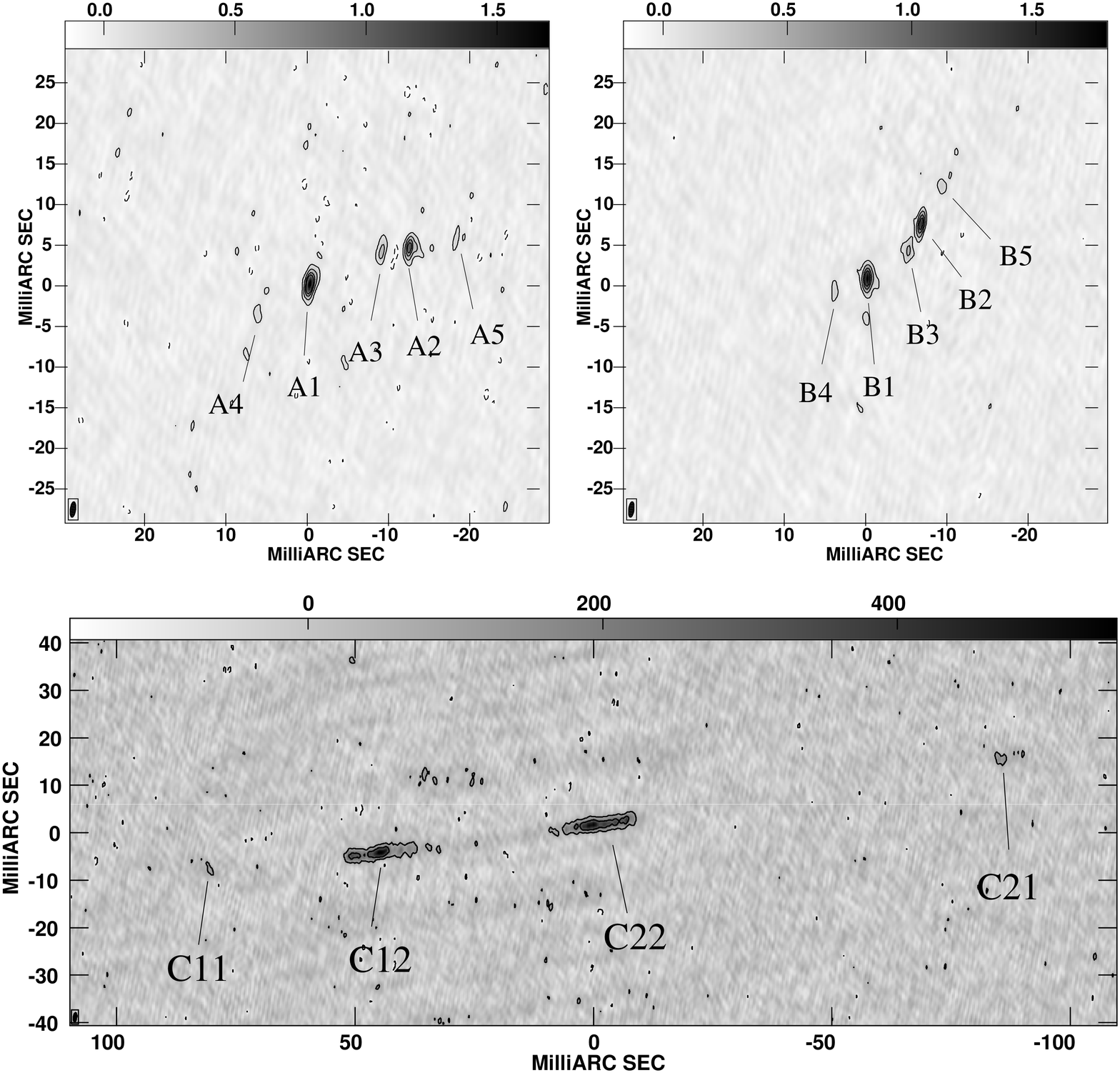}
\caption{The HSA maps of lensed images A and B, and region C at 8.4~GHz with a weighting between uniform and natural. The rms noise in the maps is 33~$\mu$Jy~beam$^{-1}$ and the contours are ($-$3, 3, 6, 12, 24, 48)~$\times$ the rms noise in the map. The size of the restoring beam is 1.9$\times$0.7~mas$^2$ and the position angle is $-$7 deg. The grey-scales are in mJy~beam$^{-1}$. North is up and east is left.}
\label{glo_3.6}
 \end{center}
\end{figure*}

Fig.~\ref{glo_1.7} shows the lensed images A and B, and the pair of merging images in region C at 1.7~GHz. At high resolution, the lensed image A shows a rich structure of five components. Image B also shows extended emission but is not so well resolved. In the earlier EVN 5~GHz observations \citep{koopmans02b}, the lensed images A and B showed two components each with a hint of substructure. Since the pair of components A1-B1 and A2-B2 have a high surface brightness, these were detected in the earlier EVN 5~GHz observations. 

In the new global VLBI 1.7~GHz data presented here, images A and B are found to have three and two new components, respectively. Elliptical Gaussian model fitting of the images was carried out using Powell's minimization routine \citep{press92} and the results are given in Table \ref{taglo1.7}. The errors at all of the frequencies were determined based on the principles described in \citet[]{fom99}. All of the components are numbered in decreasing order of their total intensity. Image A is fitted with five Gaussian components and image B is fitted with only four Gaussian components. Since image B is the counterpart of image A with a relative magnification $\sim$~1, image B is also expected to have five components. The component B3+B2 is identified as the composite of the unresolved components B3 and B2. Furthermore, if B3 has a higher peak flux density than B2 (like its counterpart A3 in image A), then the peak position of B3+B2 will be closer to the peak position of B3. Thus, the model-fitted components of image A may not represent the counterpart components of image B.

Fig.~\ref{glo_5} shows the high resolution global VLBI images at 5~GHz. Image A clearly shows the five components which were not well-resolved at 1.7~GHz. Here, image B also shows the expected five-component structure, that is, with components B3 and B2 now clearly resolved. Component 2 in images A and B has a higher surface brightness than component 3, whereas at 1.7~GHz, component 3 has a higher surface brightness than component 2 in image A. This suggests the presence of frequency dependent structure. The model fitting was again carried out with Powell's minimization routine. Images A and B were fitted with five components each, the results of which are presented in Table \ref{taglo5}.

The series of components observed in the lensed images A and B at 1.7 and 5~GHz are non-collinear which can be used to test the opposite parity expected from the theory of gravitational lensing. The opposite parity of images A and B is clearly demonstrated by the opposite curvature of the series of components.

\subsubsection{Region C}
Region C was previously known to have four components and all of these were detected in the 1.7 and 5~GHz imaging (see Fig.~\ref{glo_1.7} and \ref{glo_5}). The pair of merging images in region C are referred to as C1 (the east pair) and C2 (the west pair). The components have been numbered in an ascending order going inwards from the outside. The outer components are labelled as C11 and C21 for the east and west pair, respectively. The elongated components are labelled as C12 and C22, which are further resolved into several components. Each of these components are labelled as a, b etc. going inwards from the outside, for example, C12a, C12b, etc. in C1 and C22a, C22b, etc. in C2 (see Fig.~\ref{glo_5}). Note that this labelling convention is different from that used previously.

At 1.7~GHz, the outer components on either side (C11 and C21) are fitted with two Gaussian model components each. However, due to the low resolution and low signal-to-noise ratio (SNR), these are identified as a single component each, and their centroid (flux-density weighted) positions are reported here. The inner pair of elongated components were each fitted by three Gaussian model components. The peak positions, the peak flux densities and the total intensities from the Gaussian model fitting are given in Table \ref{taglo1.7}. Note that the components a, b and c found in component C12 may not be the counterparts of C22 for three reasons. Firstly, these components are not well-resolved. Secondly, if these are a pair of merging images straddling the critical curve, then the magnification gradient is rapidly changing on either side of the critical curve. Ideally, it should change similarly on either side. However, practically, this may not be observed. Thirdly, the ratio of component separations of the west pair C2 (C21-C22) and east pair C1 (C11-C12) is asymmetric indicating different stretching on either side. 

At 5~GHz, the four components of region C are further resolved and show fine structure. Therefore, centroid positions were determined for most of the components by fitting multiple Gaussian components. Here, component C11 is fitted with one Gaussian. C12 was modeled with six Gaussian components whereas C22 was fitted with seven Gaussian components. The component C21 is extended, and hence, fitted with three unresolved components. Three centroid positions were determined for each of the elongated components C12 and C22, and one centroid position for C21. The peak positions, flux densities and total intensities for A and B are presented in Table \ref{taglo5}. The centroid positions, peak flux densities and the total flux densities are also given for region C in Table \ref{taglo5}.

\begin{table*}
\begin{center}
\caption{The positions relative to component A1, peak surface brightness and total flux densities of the components of lensed images A and B, and region C at 1.7~GHz from the global VLBI observations. The phase-referenced position of component A1 is $20^h 19^m 18.188^s + 11^{\circ} 27' 14.638''$ for the epoch J2000 at 1.7~GHz.}
\label{taglo1.7}
\begin {tabular}{l r r c c}
\hline 
  Comp. & RA  & Dec   &  $I_{peak}$ & $S_{total}$  \\
   &  (mas) & (mas)  &  (mJy~beam$^{-1}$) & (mJy) \\
\hline
  A1     &   0.00$\pm$0.02  &   0.00$\pm$0.02  & 19.7$\pm$1.6  &  24.3$\pm$2.5  \\
  A2     & $-$12.2$\pm$0.1  &   4.5$\pm$0.1  &  2.4$\pm$0.5  &  2.8$\pm$0.7  \\
  A3     &  $-$8.6$\pm$0.1  &   3.8$\pm$0.1  &  3.5$\pm$0.6  &  4.5$\pm$0.9  \\
  A4     &   6.2$\pm$0.1  &  $-$3.8$\pm$0.1  &  4.1$\pm$0.6  &  6.5$\pm$1.2  \\
  A5     & $-$18.0$\pm$0.3  &   5.5$\pm$0.3  &  1.1$\pm$0.3  &  1.7$\pm$0.6  \\
\hline                                       
  B1     &  $-$3005.74$\pm$0.02  & $-$1503.63$\pm$0.02  & 11.5$\pm$0.1  &  10.9$\pm$1.6  \\
  B3+B2  &  $-$3011.0$\pm$0.1  & $-$1498.0$\pm$0.1  &  3.1$\pm$0.6  &   2.9$\pm$0.8  \\
  B4     &  $-$3004.5$\pm$0.1  & $-$1502.6$\pm$0.1  &  8.5$\pm$1.0  &  21.9$\pm$2.8  \\
  B5     &  $-$3012.5$\pm$0.2  & $-$1495.9$\pm$0.2  &  2.5$\pm$0.5  &   4.2$\pm$1.0  \\
\hline                                        
  c11    &  $-$2013.6$\pm$0.2  & $-$3233.3$\pm$0.2  & 2.1$\pm$0.4  &  11.6$\pm$2.3  \\
  c12a   &  $-$2045.4$\pm$0.1  & $-$3229.9$\pm$0.1  & 4.1$\pm$0.6  &  15.8$\pm$2.4  \\
  c12b   &  $-$2053.2$\pm$0.3  & $-$3228.7$\pm$0.3  & 1.5$\pm$0.4  &   4.6$\pm$1.2  \\
  c12c   &  $-$2061.0$\pm$0.3  & $-$3228.2$\pm$0.3  & 1.5$\pm$0.4  &   5.7$\pm$1.4  \\
\hline                                       
  c21    &  $-$2178.8$\pm$0.1  & $-$3209.6$\pm$0.1  &  4.9$\pm$0.7  &  18.4$\pm$2.1  \\ 
  c22a   &  $-$2098.3$\pm$0.2  & $-$3221.3$\pm$0.2  &  1.7$\pm$0.4  &   7.2$\pm$1.7  \\
  c22b   &  $-$2092.2$\pm$0.3  & $-$3223.9$\pm$0.3  &  3.1$\pm$0.5  &  17.5$\pm$3.0  \\
  c22c   &  $-$2083.5$\pm$0.6  & $-$3227.3$\pm$0.6  &  0.8$\pm$0.3  &   3.3$\pm$1.2  \\
               
  \hline
\end {tabular}
\end{center}
\end{table*}

\begin{table*}
\begin{center}
\caption{The positions relative to component A1, peak surface brightness and total flux densities of the components of lensed images A and B, and region C at 5~GHz from the global VLBI observations. For all of region C, except C11, the centroid positions and the total flux densities are given. The phase-referenced position of component A1 is $20^h 19^m 18.187^s + 11^{\circ} 27' 14.635''$ for the epoch J2000 at 5~GHz. }
\label{taglo5}
\begin {tabular}{l r r c c}
\hline
  Comp. & RA  & Dec   &  $I_{peak}$ & $S_{total}$  \\
   &  (mas) & (mas)  &  (mJy~beam$^{-1}$) & (mJy) \\
\hline
  A1   &   0.00$\pm$0.02 &   0.00$\pm$0.02  & 4.4$\pm$0.6 & 6.5$\pm$1.0  \\
  A2   & $-$12.20$\pm$0.02 &   4.44$\pm$0.02  & 3.3$\pm$0.5 & 3.6$\pm$0.7  \\
  A3   &  $-$8.5$\pm$0.1 &   3.7$\pm$0.1  & 0.8$\pm$0.2 & 1.8$\pm$0.6  \\
  A4   &   6.6$\pm$0.1 &  $-$3.7$\pm$0.1  & 0.6$\pm$0.2 & 0.9$\pm$0.4  \\
  A5   & $-$17.4$\pm$0.3 &   5.7$\pm$0.3  & 0.4$\pm$0.2 & 1.6$\pm$0.7  \\
\hline                                 
  B1   & $-$3005.95$\pm$0.02 & $-$1503.94$\pm$0.02  & 4.9$\pm$0.6 & 7.1$\pm$1.1  \\
  B2   & $-$3012.48$\pm$0.02 & $-$1497.20$\pm$0.02  & 3.2$\pm$0.5 & 3.0$\pm$0.6  \\
  B3   & $-$3011.5$\pm$0.1 & $-$1499.4$\pm$0.1  & 1.0$\pm$0.3 & 2.2$\pm$0.6  \\
  B4   & $-$3002.0$\pm$0.1 & $-$1505.7$\pm$0.2  & 0.7$\pm$0.2 & 1.5$\pm$0.5  \\
  B5   & $-$3015.1$\pm$0.2 & $-$1492.7$\pm$0.2  & 0.4$\pm$0.2 & 0.7$\pm$0.3  \\
\hline                                 
  c11    &  $-$2012.3$\pm$0.2   & $-$3234.2$\pm$0.2   & 0.8$\pm$0.2   &  4.7$\pm$1.4 \\
  c12a   &  $-$2043.8$\pm$1.0   & $-$3231.0$\pm$1.0   & 1.5$\pm$0.3   &  4.8$\pm$1.1  \\
  c12b   &  $-$2049.1$\pm$1.0   & $-$3230.3$\pm$1.0   & 1.7$\pm$0.3   &  9.1$\pm$0.3  \\
  c12c   &  $-$2056.3$\pm$1.0   & $-$3229.6$\pm$1.0   & 0.7$\pm$0.2   &  4.6$\pm$1.1  \\
\hline                                       
  c21    &  $-$2179.4$\pm$0.1   & $-$3210.4$\pm$0.1   & 0.9$\pm$0.2   &  5.2$\pm$1.0 \\ 
  c22a   &  $-$2097.6$\pm$1.0   & $-$3210.4$\pm$1.0   & 1.3$\pm$0.3   &  8.0$\pm$1.4  \\
  c22b   &  $-$2091.3$\pm$1.0   & $-$3224.9$\pm$1.0   & 1.5$\pm$0.3   &  7.3$\pm$1.3  \\
  c22c   &  $-$2086.2$\pm$1.0   & $-$3225.5$\pm$1.0   & 0.9$\pm$0.2   &  5.0$\pm$0.9  \\
 \hline
\end {tabular}
\end{center}
\end{table*}

\subsection{High Sensitivity Array (HSA)}

\subsubsection{Observations and data reduction}

Further high resolution observations at 8.4~GHz were needed to independently confirm the series of components detected in the lensed images at 5~GHz, and to better determine the spectra of lensed images A and B. Since the background quasar has a steeply falling radio spectrum ($\alpha_{1.7}^{5} = -$0.94, where $S_{\nu} \propto \nu^\alpha$), observations at a higher frequency demanded increased sensitivity. Moreover, to carry out a spectral analysis of the finely resolved structure, high frequency and high resolution imaging was needed. Therefore, MG~2016+112 was observed with the HSA at 8.4~GHz. The HSA included the following large antennas: the 305m$-$Ar, 100m$-$Eb, 100m$-$Green Bank Telescope and phased VLA, in addition to the 10 VLBA antennas.

The observations were made on 2006 April 30 and lasted for 7.5~h. The right hand and left hand circular polarisation data were recorded together in four IFs, each with 8~MHz bandwidth and 16 channels. No cross polarisation was performed. The data were correlated with the VLBA correlator using an integration time of 1~s to reduce time averaged smearing (further observational details can be found in Table \ref{obs_scan}). Ar had a power failure and problems with software which allowed observations for only 1.5~h from the 3~h window available on-source. The data were reduced by following a similar technique to that used in reducing the global 1.7 and 5~GHz data. The images were weighted to obtain an optimum combination of sensitivity and resolution in the maps as described in Sect. \ref{20mer}. 

\begin{table*}
\begin{center}
\caption{The positions relative to A1, peak surface brightness and total flux densities of the components of lensed images A and B, and region C at 8.4~GHz from the HSA observations. The phase-referenced position of component A1 is $20^h 19^m 18.187^s + 11^{\circ} 27' 14.634''$ for the epoch J2000 at 8.4~GHz.}
\label{taglo8.4}
\begin {tabular}{l r r c c}
\hline
  Comp. & RA  & Dec   &  $I_{peak}$ & $S_{total}$  \\
             &  (mas)      & (mas)       &  (mJy~beam$^{-1}$) & (mJy) \\ \hline
  A1   &   0.0$\pm$0.1 &   0.0$\pm$0.1  & 1.6$\pm$0.1 & 2.6$\pm$0.1  \\
  A2   & $-$12.3$\pm$0.1 &   4.5$\pm$0.1  & 0.9$\pm$0.1 & 1.4$\pm$0.1  \\
  A3   &  $-$8.8$\pm$0.1 &   3.9$\pm$0.1  & 0.3$\pm$0.1 & 0.5$\pm$0.1  \\
  A4   &   6.4$\pm$0.4 &  $-$3.7$\pm$0.4  & 0.2$\pm$0.1 & 0.3$\pm$0.1  \\
  A5   & $-$17.9$\pm$0.2 &   5.3$\pm$0.2  & 0.1$\pm$0.1 & 0.1$\pm$0.1  \\
\hline                                 
  B1   & $-$3005.5$\pm$0.1 & $-$1503.7$\pm$0.1  & 1.7$\pm$0.1 & 2.8$\pm$0.1  \\
  B2   & $-$3012.1$\pm$0.1 & $-$1497.0$\pm$0.1  & 1.4$\pm$0.1 & 1.8$\pm$0.1  \\
  B3   & $-$3010.6$\pm$0.4 & $-$1500.3$\pm$0.4  & 0.4$\pm$0.1 & 0.6$\pm$0.1  \\
  B4   & $-$3001.5$\pm$0.2 & $-$1505.9$\pm$0.2  & 0.1$\pm$0.1 & 0.3$\pm$0.1  \\
  B5   & $-$3014.6$\pm$0.2 & $-$1492.5$\pm$0.2  & 0.2$\pm$0.1 & 0.3$\pm$0.1  \\
\hline                                 
  c11  & $-$2012.3$\pm$0.3 & $-$3233.3$\pm$0.3  & 0.2$\pm$0.1 & 0.4$\pm$0.1 \\
  c12  & $-$2047.8$\pm$0.3 & $-$3229.6$\pm$0.3  & 0.6$\pm$0.1 & 7.5$\pm$0.7 \\
 \hline                                 
  c21  & $-$2177.8$\pm$0.5 & $-$3209.9$\pm$0.5  & 0.2$\pm$0.1 & 1.2$\pm$0.3 \\ 
  c22  & $-$2092.8$\pm$0.3 & $-$3223.7$\pm$0.3  & 0.5$\pm$0.1 & 8.5$\pm$1.0 \\
  \hline
\end {tabular}
\end{center}
\end{table*}
\subsubsection{Results}
The images of A, B and region C are shown in Fig.~\ref{glo_3.6}. The HSA imaging at 8.4~GHz confirmed the detection of three new components in images A and B, in addition to the two components known from the earlier observations. The model fitting of the images A and B was done using the \textsc{aips} task \textsc{jmfit}. A and B were both fitted with 5 Gaussian components each. The lensed images of MG~2016+112 are weak at 8.4~GHz (mJy level). Hence, the faint components 4 and 5 in images A and B and the pair of merging images in C have a low SNR. The total flux densities of these components were therefore measured by integrating the surface brightness within the 3~$\sigma$ confidence boundary (where $\sigma$ is the rms map noise). The positions of these components were found from the peak of the surface brightness. The peak surface brightness, the integrated flux densities and the peak positions of the components are presented in Table \ref{taglo8.4}. The errors were determined using the same method as before.

\label{rad_spec16}
\begin{figure*}
\begin{center}
  \includegraphics[scale=0.3]{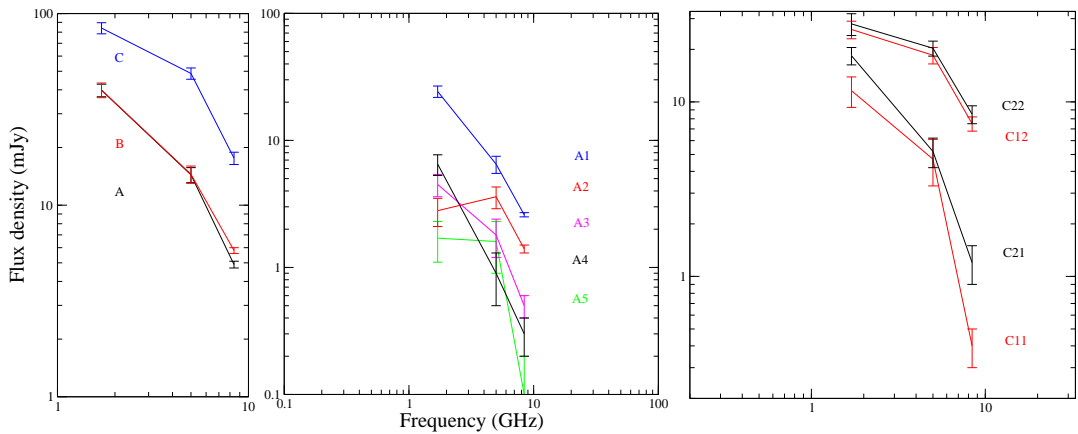}
\caption{{\it (left)} The radio spectra of lensed images A and B, and region C. {\it (centre)} The radio spectra of the components of lensed image A. {\it (right)} The radio spectra of the components in region C. The data are taken from the global VLBI and HSA observations between 1.7 and 8.4 GHz.}
\label{20spec}
 \end{center}
\end{figure*}

\subsection{Emission from a fifth lensed image or the lensing galaxy}
To estimate an upper limit on the flux density of a possible fifth image located in the vicinity of the lensing galaxy D, a fourth sub-field was mapped along with the sub-fields centred on images A, B and C. The sub-field centred on the optical position of the primary lens D at 1.7, 5 and 8.4~GHz was \textsc{clean}ed using the task \textsc{imagr}. The sizes of the fields mapped were 0.51$\times$0.51~arcsec  at 1.7~GHz, 0.3$\times$0.3~arcsec at 5~GHz and a field of size 0.2$\times$0.2~arcsec at 8.4~GHz. The fields were naturally weighted in order to achieve maximum sensitivity. However, no radio emission from a fifth component or from the lensing galaxy was found. The flux density limits (5$\sigma$-level) are 0.41, 0.18 and 0.10~mJy~beam$^{-1}$ at 1.7, 5 and 8.4~GHz, respectively. From lens models with a small constant-density core-radius, a highly demagnified image is predicted \citep{narasimha86}. Thus, the fifth lensed image is not expected to be detected in the observations presented here.

\section{Radio spectral energy distributions}
\label{spec}
Using the data from the high resolution radio observations of the lensed images, it is now possible to carry out a spectral analysis of the lensed radio components. In Fig.~\ref{20spec}, the integrated spectral energy distributions of the two lensed images A and B, and of region C between 1.7 and 8.4~GHz is shown. Images A and B have similar radio spectra and flux densities, as expected from gravitational lensing, and the spectra are consistent with previous multi-epoch and multi-frequency observations. However, there is evidence for lensed image A having a lower flux-density than image B at 8.4~GHz. This is probably due to image A being more resolved at this frequency compared to image B (see from Fig. \ref{glo_3.6} that image A is extended in an east-west direction, which has the best angular resolution). The radio spectrum of region C is found to be slightly flatter relative to images A and B. Furthermore, the flux density of region C is significantly higher, as expected for highly magnified images near the critical curve. 

Also shown in Fig.~\ref{20spec} are the spectra of the five detected radio components of image A. Components 1, 3 and 4 are found to have steep radio spectra between 1.7 and 8.4~GHz, which is consistent with optically thin, extended radio structure. Component 5 appears to have a flatter radio spectrum between 1.7 and 5~GHz, before steepening at 8.4~GHz. However, the uncertainties in the flux-densities of component 5 are quite large, so the overall radio spectrum could be steeper. The only genuine flat-spectrum radio component is 2, which shows a turnover at 5~GHz. This optically thick radio component is therefore assumed to be the core of the radio source and coincident with the strong optical/infrared quasar emission found in the {\it HST} imaging. Thus, like the optical core, component 2 is doubly imaged.

Finally, Fig.~\ref{20spec} also shows the radio spectra of the four components in region C between 1.7 and 8.46~GHz. The merging images are expected to belong to the same part of the background quasar and hence, they should have similar spectra. The components of the inner pair (C12-C22) and the components of the outer pair (C11-C21) show similar spectra within the uncertainties. The inner pair of elongated components of C (C12-C22) have flatter spectra than the outer pair of components (C11-C21). Since the components of the inner pair also have higher flux densities, these dominate the spectrum of region C at low resolutions, for example, as found in the MERLIN imaging presented in Section \ref{20mer}. Also, there is evidence for the C2 components having a higher flux-density than their C1 counterparts, which also adds to the case for there being a gravitational perturbation to the mass distribution, for example, from the nearby satellite galaxy (see Section \ref{20ma_mod}).

\section{Mass Models}
\label{20ma_mod}
In this section, various mass models for the lens potential of MG~2016+112 are tested with increasing complexity. Given the large number of observational constraints and the uncertainty of the region C components, different combinations of observational constraints are also tested. The aim is to find the simplest scenario for the mass distribution and the background source that will fit the positions of the five radio components observed in images A and B, the flux density ratio of (at least) component A1 and B1, and the positions of the four components in region C. Firstly, the predictions of the K02 mass model are tested for its consistency with the new observations. Next, two-galaxy mass models are studied which take into account all of the observed constraints and the dwarf companion to the lens. Lastly, three-galaxy models are also tested and the predictions of various mass models are further discussed.

\begin{figure*}
\begin{center}
   \includegraphics[scale=0.8]{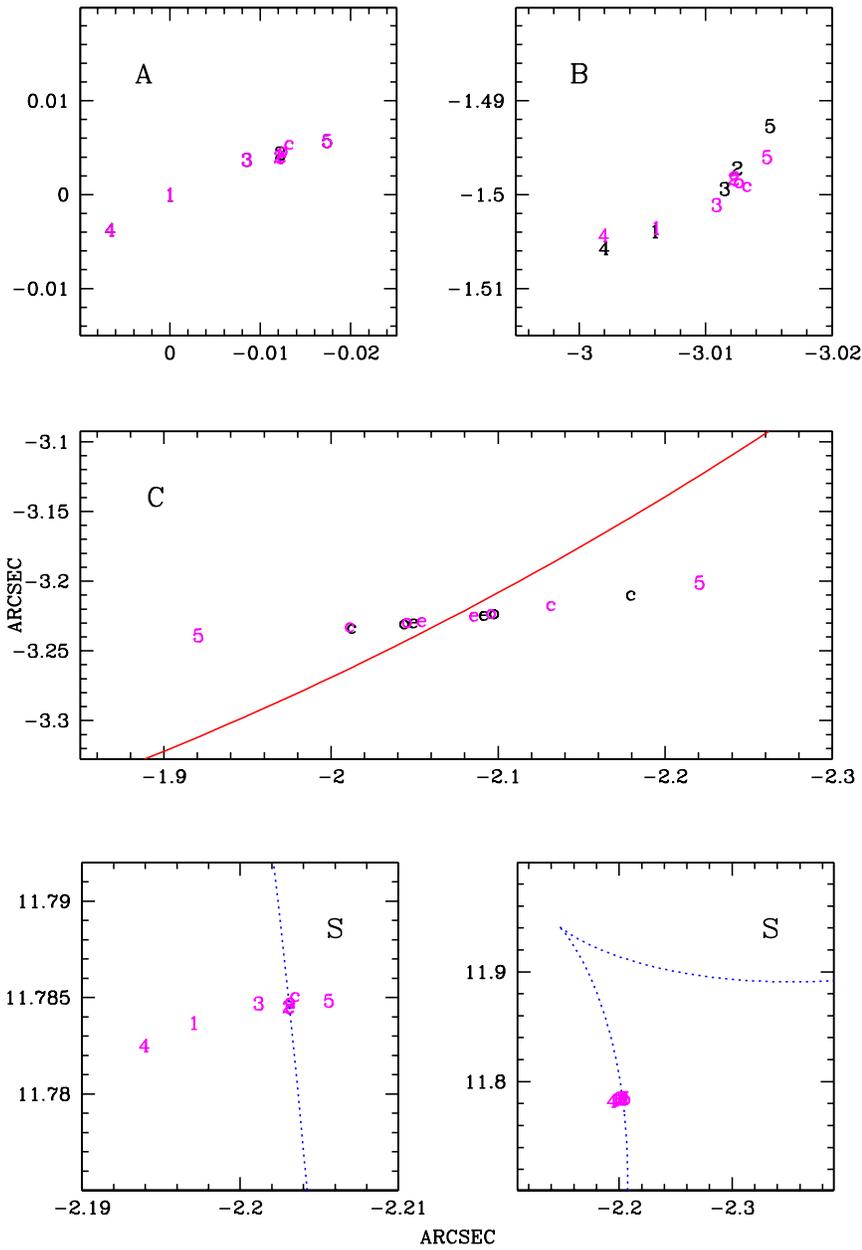}
\caption{A comparison of the observed positions (black) of the lensed components
in images A and B, and region C with those predicted (magenta) from the K02 mass
model. The critical curve in the image plane (red) is shown to split region C.
The panel S on the left shows the source positions, relative to the tangential caustic
(blue) whereas the same region at larger scales is shown in the panel to the
right. The axes are in arcsec, the offsets are relative to component A1. North is up and east is left.}
\label{kscimg}
 \end{center}
\end{figure*}

\subsection{The K02 mass model}
The multi-frequency high resolution observations were conducted with the aim of testing the K02 mass model. As a sanity check, the K02 mass model was first reconstructed from the K02 constraints. Note, however, that the K02 constraints did not include the outermost pair of components in region C (C11 and C21). Hence, the astrometric anomaly issue was not addressed in K02. In \citealt{koopmans02b}, the source component 2 was shown to be situated on a caustic. The positions of A1-B1 as doubly imaged and A2-B2-C12b-C22b as quadruply imaged components along with the flux density ratio (S$_B$/S$_A$) were used. The mass model consisted of the main galaxy D (singular isothermal ellipsoid, hereafter SIE), a mass distribution M1 (singular isothermal sphere, hereafter SIS) which contributes to the convergence coming from the environment as detected in the weak lensing analysis of \citet{clowe01}, and a mass distribution M2 (SIS) due to another physically nearby over-density of galaxies found spectroscopically by \citet{soucail01}. The mass modeling analyses carried out here were done using the publicly available code \textsc{gravlens} \citep{keeton01}. 

In light of the rich core-jet structure found from the high resolution observations presented in this paper, the reconstructed mass model of K02 was used to predict the positions and flux-densities of the different components in the lensed images. This was done using the 5~GHz global VLBI data from Table~\ref{taglo5}. The observed and predicted image positions for all of the components in lensed images A and B, and in region C are shown in Fig.~\ref{kscimg}. Here, the newly found components 3, 4 and 5 of image A were used to predict the positions of their counterparts. We find from this model that the components 4, 1, 3 and part of 2 are doubly imaged whereas component 5 and part of component 2 (c, o and e) are quadruply imaged. The labels c, o and e correspond to the merging pair of components C11-C21, C12a-C22a and C12b-C22b respectively\footnote{The labels c, o and e have no significance and are merely used for labelling convenience.}. 
An interesting result of this model is that component 5 is predicted to have four images, which are referred to as A5, B5, C1-5 and C2-5. Components C1-5 and C2-5 have opposite parity and are predicted to be about 100~mas on either side of region C (see Fig. \ref{kscimg}). Their magnification relative to A5 (or B5)\footnote{Since the flux density ratio of image A and B is $\sim$~1 and the same is almost true between C1 and C2, the components of A and C2 will be taken as the representative components in the discussion about magnification for the sake of simplicity.} is larger by a factor of $\sim$ 10. Since gravitational lensing conserves the surface brightness of the lensed images, the counterparts of A5 and B5 in region C are expected to be 10 times larger in solid angle.

\begin{figure}
\begin{center}
  \includegraphics[scale=0.35]{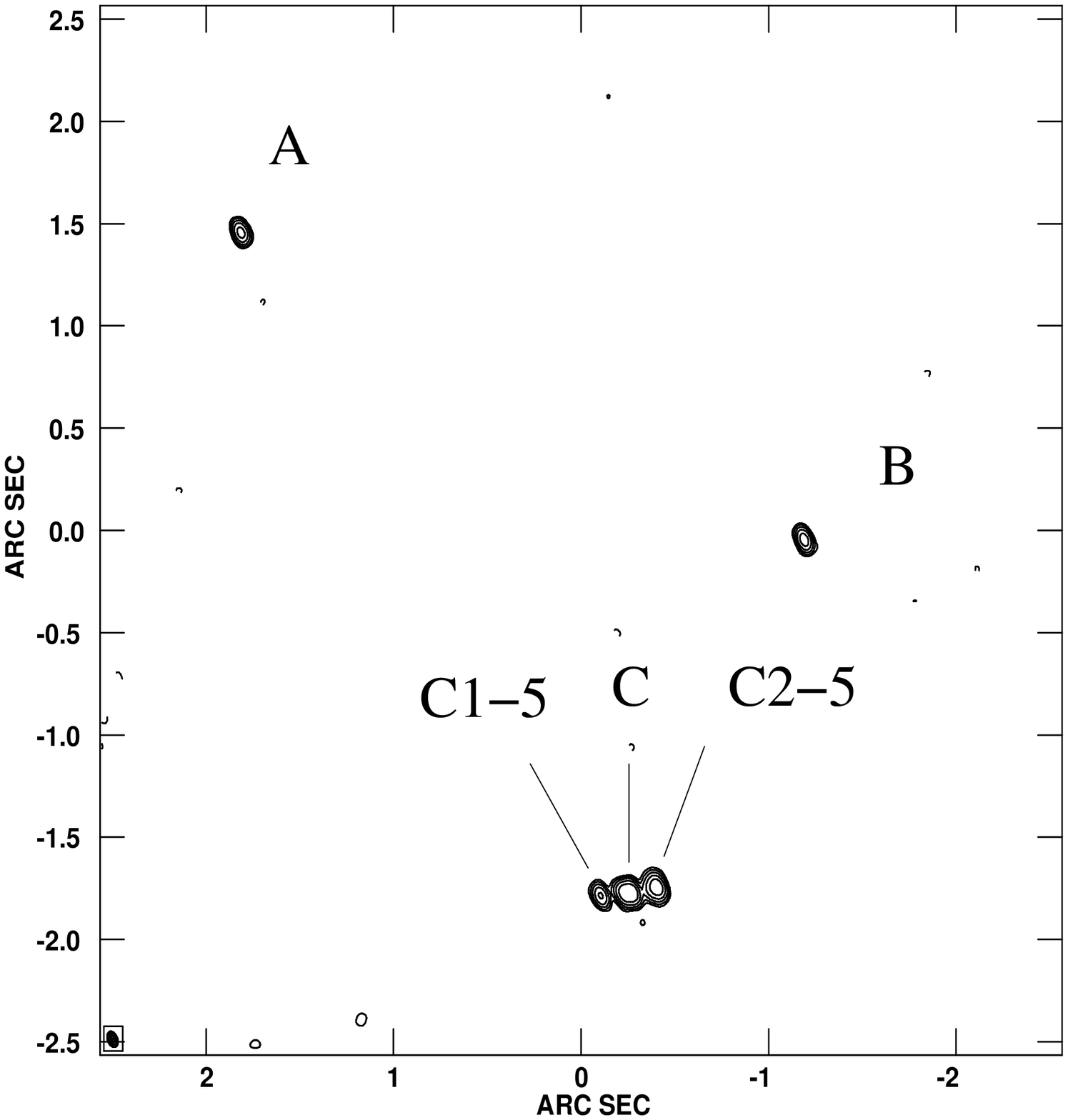}
\caption{MERLIN 5~GHz image with component 5 artificially added to the {\it uv}-data. The image has been restored with a beam of size 0.07$\times$0.05~arcsec$^2$ and position angle of 31 deg.}
\label{m6fk}
 \end{center}
\end{figure}

The detection of component 5 in region C is thus a crucial test of the K02 model. However, there is no evidence of emission near the expected position of component 5 in region C for any of the datasets presented here. In order to test that our MERLIN and VLBI datasets had sufficient resolution and surface brightness sensitivity to detect a possible fifth component in region C, we carried out simulations where artificial components were added to the $uv$-dataset and new maps were created. The MERLIN 5~GHz dataset was found to be capable of detecting component 5 (see Fig. \ref{m6fk}). The non-detection of component 5 in region C in the actual MERLIN 5~GHz map (c.f. Fig.~\ref{mer_186}) indicates that component 5 is not quadruply imaged and the K02 scenario is either incorrect, or that the surface brightness of component 5 in region C is lower than what is predicted, and the K02 scenario needs to be modified to take this into account.

It is noted that irrespective of the newly found discrepancy of the component 5 predictions with the K02 mass model, the K02 mass model was originally not complete since it did not fit the positions of the outer components (C11 and C21) on either side of region C. Also, from Fig. \ref{kscimg}, it is clear that the observed positions of the three new components found here (component 3, 4, and 5) are not reproduced in the lensed images A and B by the K02 mass model.
\subsection{Constraints and priors}

The constraints on the positions of the lensed images are taken from the high resolution global VLBI data at 5~GHz (see Table~\ref{taglo5}). The mass models will not be sensitive to an astrometric shift of the components which are $\le$~1~mas \citep{kochanek04}. Thus, the astrometric uncertainties of all components are set to a minimum of 1~mas, in spite of the higher precision offered by the VLBI observations. For all mass models, the flux density ratio S$_{B1}$/S$_{A1}$ is constrained to 1.09$\pm$0.22 (Table~\ref{taglo5}).

The lensed images A and B show opposite parity and the same five components. Therefore, all five components are at least doubly imaged. The components in region C appear to consist of two opposite parity merging images with each showing a pair of compact and extended source components. The optical emission line spectra confirms that region C is at the same redshift as the lensed images A and B. Therefore, it is unlikely that the components of region C correspond to a source other than the lensed quasar (images A and B), which happen to lie close enough to give the expected four-image configuration of a single object. Hence, region C is almost certainly related to the same background source, but it is not clear to which part observed in image A (or B), or anything that is unseen in image A (or B).

The main lensing galaxy D and the other line-of-sight galaxies are detected in the optical and their positions, measured with respect to the compact optical emission from image A, are taken from the CASTLES database. As radio component 2 in the lensed images A and B shows the flattest radio spectrum of all the detected components, it is assumed to be the radio core and coincident with the optical quasar emission (Fig. \ref{20spec}). The ellipticity of the lens galaxy D in the optical is 0.43$\pm$0.01 with a position angle of $-$59 deg east of north. Since galaxy D is a giant elliptical with a stellar velocity dispersion of $\sim$~328~km\,s$^{-1}$, it is expected to have the largest contribution to the image splitting and the ellipticity of the halo. Thus, we apply a prior on the ellipticity and fix the position angle from the values obtained from the optical surface brightness profile of galaxy D. This is to make sure that the $\chi^2$ minimization does not converge to any unreasonable mass models with an otherwise lower $\chi^2$ as compared to a desired model. Finally, from the combined lensing and stellar dynamics analysis conducted by \citet{treu02}, the logarithmic slope of the density profile of the lensing galaxy was found to be $\gamma'=$~2.0$\pm$0.1. Therefore, only elliptical isothermal mass profiles for galaxy D were considered.

\begin{figure*}
\begin{center}
  \includegraphics[scale=0.25]{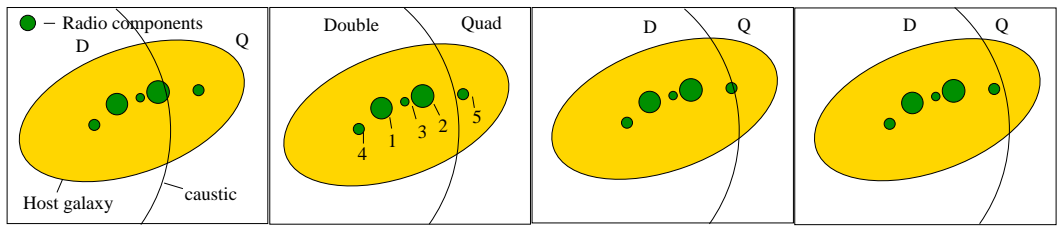}
\caption{The various scenarios A, B, C and D (from left to right) for the critical curve splitting the background source. As the critical curve move left to right, the multiplicities of the components detected in the high resolution VLBI imaging will change.}
\label{mo_scenall}
 \end{center}
\end{figure*}

\subsection {A two-galaxy model}
\label{2galmod}

We first test a two galaxy model for the mass distribution that includes the
dominant elliptical galaxy D (SIE) and the nearby dwarf galaxy G1 (SIS), which
is spectroscopically confirmed to be at the same redshift as galaxy D. The
offsets of galaxies D and G1 from the radio component A2 (assumed to be
coincident with the optical emission from image A, see Fig. \ref{mer_186}) are $\Delta$RA~$=-$1740$\pm$3~mas, $\Delta$Dec~$=-$1782$\pm$3~mas, and $\Delta$RA~$=-$2499$\pm$28~mas, $\Delta$Dec~$=-$4037$\pm$28~mas, respectively. Including the satellite G1 is essential because of its proximity to region C, which shows the asymmetry in the separated pair of opposite parity features. An external shear is also included in the mass model.  

We show different scenarios of the background source straddling the tangential caustic in the Fig.~\ref{mo_scenall} which are investigated for the two-galaxy model. The four possible scenarios are, Scenario A - the caustic goes through source component 2, Scenario B - the caustic goes between source components 2 and 5 such that it grazes source component 5, Scenario C - the caustic goes through source component 5, and Scenario D - the caustic is situated beyond source component 5. The results of the mass modeling for each scenario are summarized in Table~\ref{ta_mamod20}.

\subsubsection{Scenario A}
\textit{Constraints}: In this scenario, the caustic goes through source component 2 as was also the case for the K02 scenario\footnote{Note that scenario A and the scenario of K02 are the same. The difference lies in the data constraints that were used to make the mass models. Hence, they predict slightly different image positions and magnifications.}. The source components 4, 1, 3 and a part of 2 fall in the doubly imaged region, whereas a part of component 2 and the whole of component 5 are quadruply imaged (see Fig.~\ref{mo_scenall}). The inner components of C (i.e., labelled - o and e) are chosen as the counterparts of A2 and an unseen A2e to the west, respectively, while the outer components (i.e., labelled - c) are associated with some unseen component A2c. The uncertainties for all observed components are chosen to be 1~mas, whereas for the unseen components they are chosen to be 5~mas so that they do not contribute significantly to the $\chi^2$. Exceptions to this are the uncertainties of the counterparts of component A5 in region C, that is, C1-5 and C2-5. Their uncertainties are chosen to be very high (i.e., 10$^4$~mas) since their positions might be affected due to substructure in an unexpected way. A total of 41 constraints exist for this mass model. There are 14 free source positions and four free model parameters, that is, the Einstein radii of the two galaxies, the external shear and its position angle. The degrees of freedom (dof) are 23.

~

\begin{figure}
\begin{center}
   \includegraphics[scale=0.65]{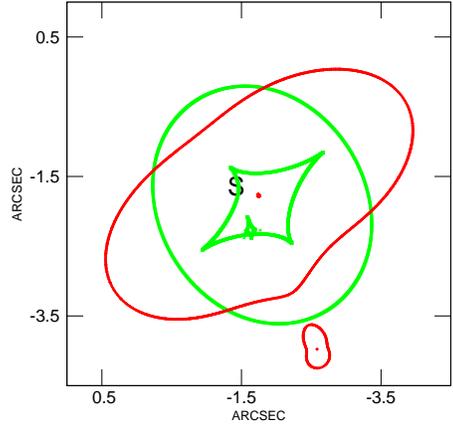}
\caption{The critical curves (red) and the caustics (green) for the two-galaxy
lens model. The position of the source (S) relative to the tangential caustic is
shown. The x and y axes show separations relative to component A1
which is chosen as the origin throughout.}
\label{sc4cc}
\end{center}
\end{figure}

\begin{figure*}
\begin{center}
   \includegraphics[scale=0.6]{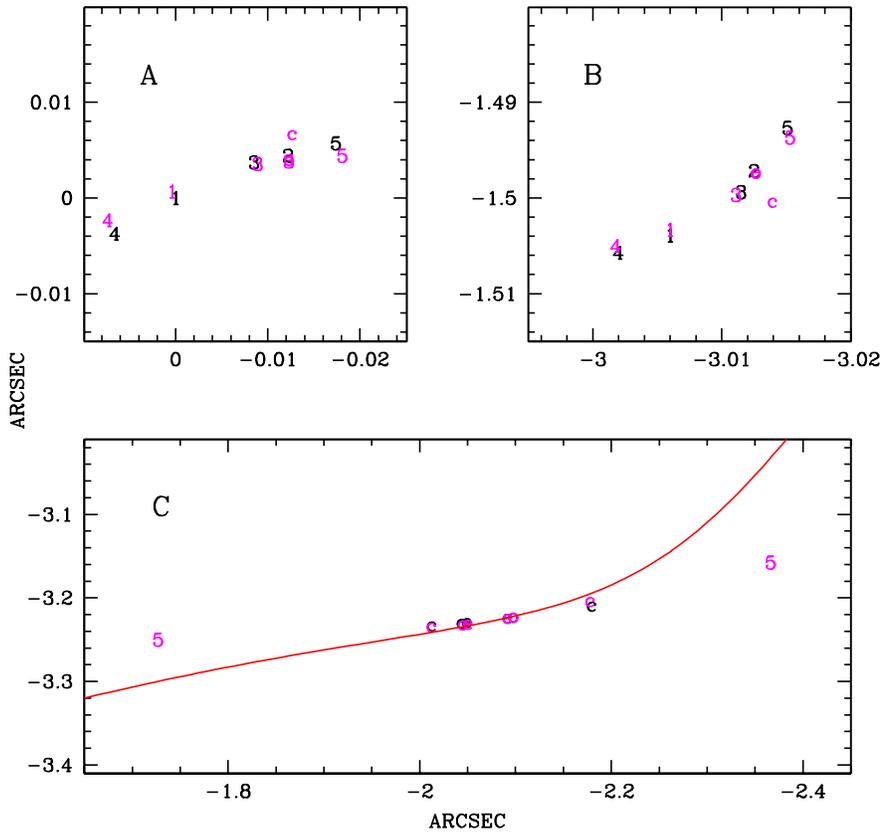}
\caption{The observed (black) and predicted (magenta) positions of the components in lensed images A and B, and region C for scenario A. The critical curve is shown in red. The axes are in arcsec and the offsets relative to component A1. North is up and east is left.} 
\label{mo_sc12}
\end{center}
\end{figure*}

\noindent \textit{Results/Predictions}: The best-fitting model has a reduced $\chi^2 \sim$~3.5. The largest contribution to the total $\chi^2$ comes from the image positions ($\chi^2 \sim$ 60, most of which is from the components of region C) and the second largest is from the galaxy positions (total $\chi^2 \sim$ 18; most of which is due to the satellite G1). The model-predicted Einstein radii of the galaxies D and G1 were found to be 1.570 and 0.143~arcsec, respectively. The recovered ellipticity of D was 0.42. The external shear was found to be 10 per cent with a position angle of $-$41.5 deg measured East of North. The critical curves and the caustics for the two-galaxy model can be seen in Fig.~\ref{sc4cc}. The observed and fitted image positions are shown in the top and bottom panels of Fig.~\ref{mo_sc12}. The relative magnification of the components in region C, with respect to A2--A2o, is very high. For example, S$_{e}$/S$_{A2e} \sim$~10$^3$ and S$_{c}$/S$_{A2c} \sim$~10$^2$. Hence, the counterparts of region C in images A and B are predicted to be 100--1000 times fainter (at a level of $\sim$10~$\mu$Jy) and 10--30 times smaller in size. These counterparts can not be detected with the sensitivity of the observations undertaken at any of the three frequencies presented here. However, component 5 is predicted to have two counter-images in region C which are referred to as C1-5 and C2-5. The predicted flux density ratio is S$_{C2-5}$/S$_{A5} \sim$~15, and a size that is at least four times larger than that of A5. Since C1-5 and C2-5 are not detected in the MERLIN observations, this scenario is ruled out.

\subsubsection{Scenario B}
\textit{Constraints}: In this scenario, although the caustic is situated between component 2 and 5, it is closer to component 5 such that C11-C21-A5-B5 are its four lensed images whereas the inner elongated pair of components o and e are associated with unseen components A5o and A5e to the east of component A5. Thus, components 4, 1, 3 and 2 will be doubly imaged as shown in the second panel of Fig.~\ref{mo_scenall}. The number of constraints are 41 and the free parameters remain the same, thereby, giving a dof of 23.

~

\begin{figure*}
\begin{center}
   \includegraphics[scale=0.6]{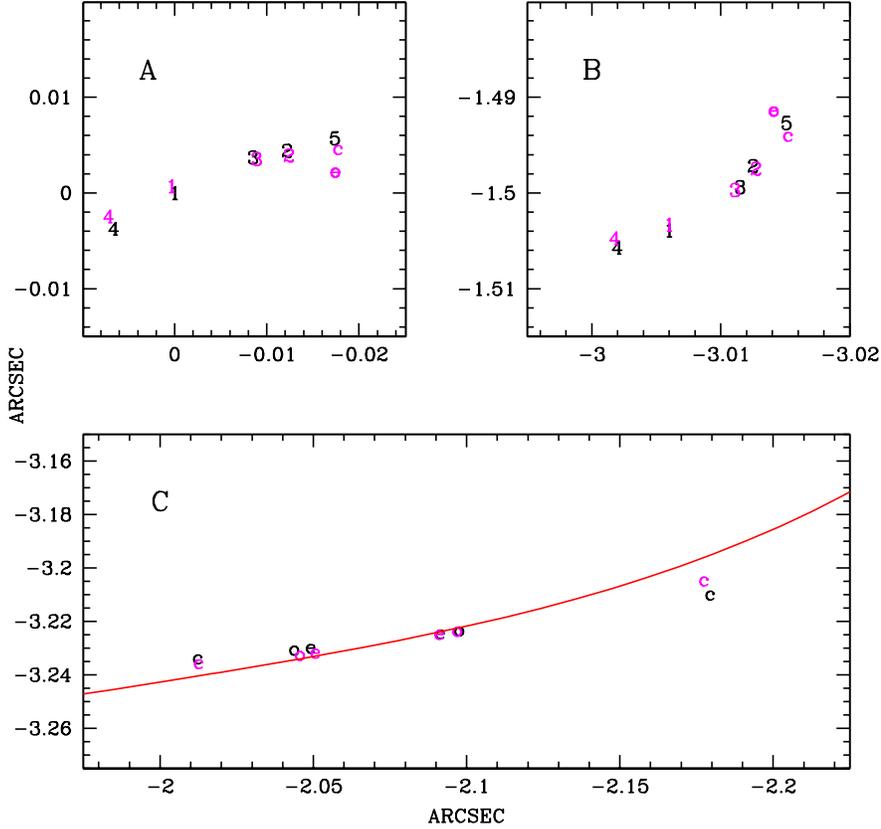}
\caption{The observed (black) and predicted positions of the components in lensed images A and B, and region C for scenario B. The critical curve is shown in red. The axes are in arcsec and the offsets are relative to component A1. North is up and east is left.}
\label{mo_sc12_B}
\end{center}
\end{figure*}

\noindent \textit{Results/Predictions}: Not surprisingly, the reduced $\chi^2$ is 3.6 because this scenario is a slight modification of scenario A and is not expected to modify the global mass model significantly. Furthermore, the individual $\chi^2$ contributions and the best-fitting model parameters are also similar. However, the predictions are expected to change here. The positions of the components in images A and B are fitted within 1~mas except for components 4 and 5, which are fitted within 1.5~mas. The observed and model predicted image positions are shown in the bottom panel of Fig.~\ref{mo_sc12_B}. The relative magnifications of the C components with respect to A are predicted to be S$_{e}$/S$_{A5e}$ $\sim~$10$^3$ and S$_{c}$/S$_{A5}$ $\sim$~200. Since the inner components of C (o and e) have very high magnification, their counterparts in images A and B would be unseen. This is consistent with the observations. However, the counterparts of component c at the position of A5 (or B5) are predicted to be $\sim$~100 times fainter whereas the observed relative magnification is $\le$~10. Due to this inconsistency, this scenario is not acceptable either.

\subsubsection{Scenario C}
\textit{Constraints}: Here, the caustic goes through component 5 such that the inner elongated pair (o--e) of region C are associated with A5 and an unseen component to the west (A5e) that are both quadruply imaged. The pair of components c are associated with an unseen component A5c that would lie further to the north-west. Components 4, 1, 3 and 2 are doubly imaged in this scenario. The total number of constraints and free parameters are the same as before, hence the dof are 23.

~

\noindent \textit{Results/Predictions}: The best-fitted model parameters are the same as before. The reduced $\chi^2$ is 3.6 and the individual $\chi^2$ contributions are similar to the previous scenarios. The doubly imaged components 1, 2 and 3 are fitted within the 1~mas uncertainties whereas component 4 is fitted within 1.5~mas. The counterparts of c, o and e in image A (or B) are predicted to be within 1.5~mas west of the peak position of component A5. The relative magnification of region C is predicted to be 100--1000 times higher than their counterparts in component A5 (or B5). Therefore, the corresponding components in images A and B would have flux densities of the order 5--50~$\mu$Jy, which is much lower than the rms noise level of the global VLBI 5~GHz observations. Moreover, it is certainly not possible to resolve these components since they would have sizes 10$-$30 times smaller. Nevertheless, since these predicted components could not be detected, their possible existence is not incompatible with our new observations. However, this scenario may or may not stand true with better observational constraints in the future.

\subsubsection{Scenario D}
\textit{Constraints}: In this scenario, the caustic is situated to the west of component 5 such that components 4, 1, 3, 2 and 5 are all doubly imaged and the components of C are associated with some undetected components in images A and B. The total number of constraints is 45. There are 16 free source positions and 4 free parameters and hence, the dof is 25.

~

\noindent \textit{Results/Predictions}: The reduced $\chi^2$ for the best-fitting model is $\sim$~3.4 and the model parameters are similar to those given in scenario A. The doubly imaged components 1, 2 and 3 are fitted within their uncertainties (1~mas) whereas components 4 and 5 are fitted within 1.5~mas. The c, o and e components are predicted to have four images. Their counterparts in A and B are expected to have a relative demagnification of 100--1000. Such components would not have been detected in the multi-frequency observations presented here. Therefore, this scenario is also consistent with the observations. In fact, observations that have 1000 times better sensitivity are needed to fully test this scenario.

\subsubsection{No constraints on the position of the lens mass components}
The cluster associated with the main galaxy D is believed to be a proto-cluster
which is not centrally concentrated yet. The conclusion that the cluster is not
virialized is inferred from the absence of any diffuse X-ray emission
\citep{chartas01}. Therefore, the optical position of the BCG (galaxy D) may not
be coincident with the centre of the cluster. Mass models with no constraint on
the position of the lensing galaxy were hence tested for different scenarios.
The following results are described for scenario D. Initially, the position of
the main galaxy D was allowed to be free. The best-fitting model shifted galaxy
D by $\sim$~60~mas to the west. The reduced $\chi^2$ of 1.5 was mainly improved
by better fitting the image positions. The total $\chi^2$ of the galaxy position
did not change significantly, which arises here solely due to the satellite G1.
The best-fitting parameters are similar to those obtained from the mass model
with the position of galaxy D constrained. The Einstein radii are 1.551 and
0.14~arcsec for galaxies D and G1, respectively. The fitted ellipticity is 0.43
and the external shear is 11 per cent with a position angle of $-$41.8 deg.
 
\subsection{A three-galaxy model}
In spite of the high external shear ($\sim$~10 per cent) for the two galaxy mass model (SIE+SIS), the image positions of all components could not be reproduced within their uncertainties. Given the dense environment around the lensing galaxy and a handful of line-of-sight nearby galaxies, an additional galaxy, G2 (SIS) was added to the mass model (i.e., SIE+SIS+SIS). The position of G2, relative to radio component A2 (assumed to be coincident with the optical emission from A) is $\Delta$RA~$=-$5749$\pm$10~mas and $\Delta$Dec~$=+$1759$\pm$10~mas. This three-galaxy model was tested for various scenarios. However. introducing G2 to the model did not result in any significant change in either the reduced $\chi^2$ or the best-fitted model parameters (see Table \ref{ta_mamod20} for details).
 
\begin{table*}
\begin{center}
\caption{The results from the mass modeling analysis and the fitted parameters. The $\chi_{\rm{ip}}^2$ and $\chi_{\rm{if}}^2$ denote the $\chi^2$ due to the image positions and flux densities, respectively. The $\chi_{\rm{gp}}^2$ denotes the $\chi^2$ due to the positions of all of the galaxies in a model, whereas $\chi_{\rm{oth}}^2$ denotes the $\chi^2$ contribution from other priors, for example, the ellipticity. See text for further details. }
\label{ta_mamod20}
\begin {tabular}{l c c c c c l l}\\
\hline
Model--Scenario & $\chi^2_{\rm{tot}}$  &  $\chi^2_{\rm{ip}}$ & $\chi^2_{\rm{if}}$ &  $\chi^2_{\rm{gp}}$ & $\chi^2_{\rm{oth}}$ & Model parameters & Comments/Status \\
\hline
SIE+SIS+shear & & & & & & & \\
Scenario A  & 81.0 & 60.0 & 1.9 & 18.0  & 0.3 & $b_D$=1.570,~~ $P_D$=($-$1.746, $-$1.777), & predicts extra components in region C \\
&&&&&&$e_D$=0.42, ~~ $\gamma_s$= (0.10, $-$41.5), &  -- not found in data\\
&&&&&&$b_{G1}$=0.143~~, $P_{G1}$=($-$2.572, $-$3.972) &  (Not acceptable) \\

Scenario B  & 83.4 & 60.6 & 1.9 & 20.8 & 0.05  &  $b_D$=1.568,~~ $P_D$=($-$1.746, $-$1.776),  & relative magnification of component 5 \\
&&&&&&$e_D$=0.43, ~~ $\gamma_s$= (0.10, $-$41.5), &  component C21--do not match data \\
&&&&&&$b_{G1}$=0.146,~~ $P_{G1}$=($-$2.575, $-$3.976) &  (Not acceptable) \\

Scenario C  & 83.5 & 61.5 & 1.9 & 20.0 & 0.07 &  $b_D$=1.569,~~ $P_D$=($-$1.746, $-$1.776),  & relative magnifications are consistent \\
&&&&&&$e_D$=0.43, ~~ $\gamma_s$= (0.10, $-$41.5), &  with the data presented here \\
&&&&&&$b_{G1}$=0.145,~~ $P_{G1}$=($-$2.573, $-$3.977) &  (acceptable $-$ needs further investigation) \\

Scenario D  & 85.0 & 62.6 & 1.9 & 20.3  & 0.04 &  $b_D$=1.569,~~ $P_D$=($-$1.746, $-$1.776),  & relative magnifications are consistent \\
&&&&&&$e_D$=0.43, ~~ $\gamma_s$= (0.10, $-$41.5), & with the data presented here  \\
&&&&&&$b_{G1}$=0.146,~~ $P_{G1}$=($-$2.573, $-$3.979) &  (acceptable $-$ needs further investigation) \\
\\
position of & 41.0 & 21.3 & 3.1 & 16.3 & 0.2 & $b_D$=1.551,~~ $P_D$=($-$1.799, $-$1.779), &  an offset of $\sim 60$~mas from the optical \\
Gal D free &&&&&&$e_D$=0.43, ~~ $\gamma_s$= (0.11, $-$41.8), & position is significantly large \\
-- Scenario D &&&&&&$b_{G1}$=0.144,~~ $P_{G1}$=($-$2.581, $-$3.959) &  (Not acceptable) \\
\hline 
SIE+SIS+SIS+shear \\
Scenario A & 79.5 & 59.0 & 1.9 & 18.0 & 0.7 & $b_D$=1.551,~~ $P_D$=($-$1.799, $-$1.779),  & Adding G2 does not improve \\
&&&&&&$e_D$=0.42, ~~ $\gamma_s$= (0.09, $-$40.5), & the model fit to the data  \\
&&&&&&$b_{G1}$=0.146,~~ $P_{G1}$=($-$2.574, $-$3.964) &   (Not acceptable)  \\
&&&&&&$b_{G2}$=0.075,~~ $P_{G2}$=($-$5.749, 1.767) &  \\ \hline

\end {tabular}     
\end{center}       
\end{table*}

\section{Discussion}
\label{disc}

In this section, we discuss the role of the satellite galaxy and the abundance of substructure in MG~2016+112. Furthermore, we compare our results for the substructure with the expectations of standard $\Lambda$CDM cosmological simulations.

\subsection{The satellite galaxy G1}
The mass distribution in MG~2016+112 definitely consists of multiple components. With a smooth mass model of SIE+shear centred at D, it is possible to fit the positions of the components only in images A and B. Fitting the components of region C results in a very poor fit with a reduced $\chi^2$ of about 180. Including the satellite galaxy (G1), significantly improves the fit to all of the astrometric constraints including the astrometric anomaly found in region C (reduced $\chi^2 =$~3.5).  Such a model allows us to estimate the mass of the satellite galaxy.

For the SIE+SIS+shear models, the Einstein radius of the satellite galaxy (i.e., the SIS component) is found to be $\sim$~0.14~arcsec. Hence, the mass of the satellite (within its Einstein radius) is found to be 8$\times 10^{9}~$M$_{\odot}$ whereas the mass of the primary lens D (within its Einstein radius) is 10$^{12}~$M$_{\odot}$ (see Eqs.~\ref{crtrad} and \ref{mcrad}). Thus, the model predicts the fraction of the mass in the satellite galaxy to be 0.8 per cent. This is consistent with the satellite mass fraction (1 per cent) suggested by \citet{kochanek04}. The model-predicted velocity dispersion of the satellite galaxy is 99~km\,s$^{-1}$. This is consistent with the mass modeling results of \citet{chen07}, who find the velocity dispersion to be 87~$< \sigma <$~101~km\,s$^{-1}$ for the satellite G1.

\subsection{Abundance of subhaloes}
In this section, we test the predictions of substructure from CDM simulations vis-{\'a}-vis the luminous substructure detected in MG~2016+112. First, we determine the number of subhaloes with a mass greater than or equal to that of the satellite galaxy expected from simulations. Second, we compare the substructure mass fraction at the projected separation of the satellite galaxy with that expected from simulations. 

For the first test, we use the results of \citet{gao04} who studied the subhalo populations over a wide range in halo mass using high resolution simulations in a $\Lambda$CDM universe. They found that the mass function of subhaloes per unit halo mass can be expressed as
\begin{equation}
\label{gaosub}
\frac{ {\rm d}N }{{\rm d} m} \sim 10^{-3.2}(m_{\rm{sub}} h/\rm{M}_{\odot})^{-1.9}~h~\rm{M}_{\odot}^{-1}\,.
\end{equation} 
The mass modeling results yield the total halo mass of the lens galaxy within the virial radius as $M_{\rm vir} \sim 2.1\times10^{13}~$M$_{\odot}$ (see Eq.~\ref{mvir}) and the mass of the satellite galaxy within its Einstein radius as $8\times10^{9}~$M$_{\odot}$. The number of subhaloes with mass a greater than or equal to the mass of the satellite galaxy (obtained by integrating Eq.~\ref{gaosub} above the satellite galaxy mass) is $\sim 3$. The detection of a single satellite galaxy suggests that the observations are consistent with the predictions within $2\,\sigma$. Note that the observed subhalo is strictly a lower limit.

For the second test, we use the results of \citet{mao04}, who investigate the substructure mass fraction, $f_{\rm sub}$, in haloes at low redshift from $\Lambda$CDM simulations\footnote{The results do not evolve significantly with redshift (Andrea Macci\`o, priv. comm.) .}. They estimate $f_{\rm sub}$ as a function of increasing projected distance within annuli of equal logarithmic width in projected separation. The satellite galaxy G1 in MG~2016+112 is at $\sim$0.05~$r_{\rm vir}$ of the main lensing galaxy D. The simulations predict $f_{\rm sub}=0.01\pm0.01$ at this projected separation \citep[see the bottom left panel of fig.~1 of][]{mao04}. While calculating the substructure mass, \citet{mao04} only use the mass within 0.025~$r_{\rm{vir}}$ of the subhalo centre. For a fair comparison, we calculate the mass of G1 within 0.025~$r_{\rm{vir}}$ using the modeling results. The mass in the smooth halo component within the requisite annuli is found by integrating the projected surface density (see Eqs.~\ref{fsubg1} \& \ref{fsubd}) of the main halo appropriately. Thus, the fraction calculated from lensing, $f_{\rm sub}= 0.09 $ (see Eq.~\ref{fsub}), is higher than the fraction predicted by the simulations. Nevertheless, these values are again consistent within $3 \sigma$.

It should be understood that the comparison of the mass fraction from lensing
with that of simulations is biased due to several reasons. The simulated
subhaloes are dark matter only whereas the substructure in MG~2016+112 is
luminous and hence, consists of baryons. Predictions from lensing will tend to
be biased towards the high end of the subhalo mass function because massive
subhaloes will produce easily detectable observable effects 
\citep[also see e.g.,][]{bryan08}. Furthermore, low
statistics is severely affecting the current comparison and hence, more examples
similar to MG~2016+112 and MG~0414+0534 are needed that could be meaningfully
compared to the predictions of simulations.

\subsection{CDM substructure near images A and B}
In MG~2016+112, the lensed images A and B show similar spectra. They exhibit the expected opposite parity. A local perturbation in any of the images can be found by comparing the relative magnification matrix with the flux ratios of the images. The presence of non-collinear structure in the images allows an accurate determination of the relative magnification matrix. The comparison shall further strengthen our belief in the absence of substructure close to images A and B.

The relative magnification ($\mu_r$) of the lensed images, that is, the magnification of an image (B) with respect to another image (A), is simply the ratio of individual magnifications $\mu_B$ and $\mu_A$. Ideally, the surface brightness of the lensed images is conserved, hence, the observed ratio of the flux densities of the images (denoted by S) should be equal to the ratio of their solid angles (denoted by $\omega$). This can be expressed as, 

\begin{equation}
\mu_r=\frac{\mu_B}{\mu_A}=\frac{S_B}{S_A}= \frac{\omega_B}{\omega_A}=|det(M_{ij}^{AB})|,
\end{equation}
where the relative magnification matrix ($M_{ij}^{AB}$) is defined as,
\begin{equation}
\label{rmm_eq}
\vec {B_1}=\left(\begin{array}{c} B_{1x} \\ B_{1y}  \end{array} \right)  = \left(\begin{array}{cc} M_{11} & M_{12} \\ M_{21} & M_{22} \\ \end{array}\right) \left(\begin{array}{c} A_{1x} \\ A_{1y}  \end{array} \right) = M_{ij}^{12} \vec {A_1}.
\end{equation}

Here, $\vec {A_1}$ and $\vec {B_1}$ denote vectors in images A and B respectively, and are counterparts of each other. This formulation can be applied to real gravitational lens systems \citep[e.g.,][]{garrett94b,jin03} provided the lensed images show a rich structure of non-collinear features. Note that at least two vectors are needed to constrain the four independent elements in the relative magnification matrix.

Consider a set of three components in A (A1, A2, A3) and B (B1, B2, B3). Let $\vec {A_{12}}$ ($\vec {B_{12}}$) and $\vec {A_{32}}$ ($\vec {B_{32}}$) denote the vectors originating from component A2 (B2) and ending at A1 (B1) and A3 (B3) respectively. The positions of the components of images A and B are taken from Tables \ref{taglo5}, after applying the appropriate shift of origin. The uncertainties are chosen as 0.1~mas for all positions throughout for simplicity. The determinant of the relative magnification matrix is found to be $-1.09\pm0.35$ and is in good agreement with the observed flux density ratios of the images. The error-bars were obtained by drawing 1000 realizations of the image positions which follow normal distributions centred at the original image positions and with a scatter equal to the uncertainty in the positions. The negative value of the determinant again confirms the opposite parity of image B with respect to image A. 

We further use the results of numerical simulations to determine the number of massive subhaloes near images A and B that would produce observable astrometric distortions. We use the subhalo mass function from Eq.~\ref{gaosub} to find the number of subhaloes in the mass range $10^7~$M$_{\odot}$ to $10^{10}~$M$_{\odot}$. The number of subhaloes at the projected separation of the images, thus, expected within an angular region of 100$\times$100~mas$^2$ is $\sim 10^{-2}$. The number of subhaloes within an annulus of width 100~mas at the projected separation of the images is $\sim 1$.    

\section{Conclusions}
\label{conc}

Multi-frequency high-resolution radio observations of the gravitational lens MG~2016+112 were conducted to carry out a spectral analysis and to find a mass model for the complex structure in the lensed images. Radio maps made with simultaneous MERLIN and global VLBI observations at both 1.7 and 5~GHz were presented. Subsequently, HSA observations at 8.4~GHz were undertaken to carry out a spectral study of the components at high resolution. In addition to the two previously known components in images A and B, three new components were detected in the observations presented here. The observations with the HSA proved crucial in the confirmation of the new components. A total of five components are now found in images A and B. No more new components are detected above 33~$\mu$Jy within a region of size 0.21$\times$0.21~arcsec$^2$ centred at images A and B from the HSA imaging at 8.4~GHz. A 5$\sigma$ upper limit was placed on the peak surface brightness of an odd image in the vicinity of the lens D, or radio emission from D, of 0.18~mJy~beam$^{-1}$ at 8.4~GHz.

The overall radio spectra and the flux densities of the components in A and B were found to be similar. The flux density ratio of images A and B were also found to be consistent with the determinant of the relative magnification matrix. Therefore, there is no significant substructure or any other effects that might affect the flux densities of the images. In region C, the morphology and spectra of C11-C21 and C12-C22 were found to be similar, as expected for lensed images. Furthermore, the measured flux-density of the C2 pair is $\sim$1.2 times the flux-density of the C1 pair at all frequencies, which could be due to the proximity of the satellite galaxy G1 to the C2 pair with a positive parity \citep{keeton03}. The identification of components in region C with those in image A (or B), on the basis of their spectra, cannot be done because the highly magnified components of region C correspond to extremely small regions in either the detected (4, 1, 3, 2 or 5) or undetected components of images A and B.

Several mass models with more than one mass component in a single lens plane were investigated for four scenarios. In these scenarios, the components of region C were constrained as the lensed counterparts of different parts of the components of images A and B, and the consequences of doing so were assessed. The mass models tested here with scenarios A and B predicted relative magnifications of the images that were inconsistent with the observations. The predictions of scenarios C and D were consistent with the observations presented here. The predictions of the mass models of \citet{koopmans02b} are not consistent with the new observations because of the detection of component 5. A SIE+SIS+shear model with the satellite galaxy G1 (SIS) found at the same redshift as the lensing galaxy D (SIE) improved the fit (reduced $\chi^2 =$~3.5) to the astrometric constraints significantly, as stated previously by \citet{kochanek04}. However, a model with more complexity, for example, ellipticity in the satellite galaxy which is currently observationally ill-constrained but can cause local azimuthal distortions near region C, is perhaps required.

We have compared the subhalo abundance and mass fraction in MG~2016+112 with lensing and simulations. About 1 to 2 subhaloes are expected in an annulus around the images from the CDM scenario and we detect one (luminous) subhalo centred at G1. Prima facie, this appears like an agreement which can be either confirmed or refuted with further deeper observations of the environment of this lens system. The subhalo mass fraction, assuming an isothermal profile, indicates a much higher mass in the subhalo centred at G1 than that expected from the simulated halo of \citet{mao04}. Considering the assumptions and caveats involved here, this comparison should only be taken as suggestive rather than conclusive.        

\section*{Acknowledgments}
AM would like to thank P. Schneider and O. Wucknitz for useful discussions. AM
is also thankful to E. Ros and the anonymous referee for their advice and suggestions
that improved the presentation of the figures. The VLBA is operated by the National Radio Astronomy Observatory which is a facility of the National Science Foundation operated under cooperative agreement by Associated Universities, Inc. The European VLBI Network is a joint facility of European, Chinese, South African and other radio astronomy institutes funded by their national research councils. MERLIN is a National Facility operated by the University of Manchester at Jodrell Bank Observatory on behalf of STFC. This work was supported by the European Community's Sixth Framework Marie Curie Research Training Network Programme, Contract No.  MRTN-CT-2004-505183 ``ANGLES". AM carried out this research while being a member of the International Max-Planck Research School (IMPRS) for Astronomy and Astrophysics.

\appendix
\section{Definitions and Equations}
The critical radius of an isothermal ellipsoidal lens is related to the velocity dispersion ($\sigma$) through the following equation  
\begin{equation}
\label{crtrad}
b_{\rm{SIE}} =\sqrt{\frac{2q}{1+q^2}} \, b_{\rm{SIS}} = \sqrt{\frac{2q}{1+q^2}} \, 4\pi \, \frac{\sigma^2}{c^2} \, \frac{D_{\rm{ds}}}{D_{\rm{s}}}
\end{equation}
which takes into account the ellipticity through the axis ratio q. The mass of the lens within the Einstein radius can then be calculated from
\begin{equation}
\label{mcrad}
M(<b_{\rm{SIS}}) = \frac{b_{\rm{SIS}}^2 c^2}{4~G} \, \frac{D_{\rm{d}} D_{\rm{s}}}{D_{\rm{ds}}}~.
\end{equation}
Here, $D_{\rm{ds}}$, $D_{\rm{d}}$ and $D_{\rm{s}}$ refer to the angular diameter distances between the lens and the source, the observer and the lens, and the observer and the source, respectively. 

~

{\noindent}The definition of virial radius $r_{\rm{vir}}$ and the mass $M_{\rm{vir}}$ of the halo within the virial radius \citep{mo98} is given by
\begin{eqnarray}
\label{rvir}
r_{\rm{vir}}&=& \frac{V_c}{10~H(z)}\\
\label{mvir}
M_{\rm{vir}}&=& \frac{V_c^3}{10~G~H(z)}~. 
\end{eqnarray}
where $V_c=\sqrt{2}\sigma$ is the circular velocity, $G$ is the gravitational
constant and $H(z)$ is the Hubble parameter.

~

{\noindent}The mass of satellite G1 within 0.025~$r_{\rm{vir}}$, the mass of lens galaxy D within an annulus and the satellite mass fraction are calculated using the following equations,
\begin{eqnarray}
\label{fsubg1}
M_{G1} &=& \int_{0}^{0.025~r_{\rm{vir}}} \frac{\sigma_{G1}^2 2 \pi r \rm{d}r }{2 \pi G r}\\
\label{fsubd}
M_D &=& \int_{R1}^{R2} \frac{\sigma_D^2 2 \pi r \rm{d}r }{2 \pi G r}\\
\label{fsub}
f_{sub}&=& \frac{M_{G1}}{M_D}=\frac{\sigma_{G1}^2}{\sigma_{D}^2}\frac{0.025~
r_{\rm{vir}}}{R1-R2}
\end{eqnarray}

Here, $R1$ and $R2$ correspond to the projected radial distances from the center
of the potential which bound the annulus.

\end{document}

%% file: journals.tex
\def\aj{AJ}%
\def\apj{ApJ}%
\def\apjl{ApJ}%
\def\apjs{ApJS}%
\def\aap{A\&A}%
\def\mnras{MNRAS}%
\def\nat{Nature}%